\documentclass{aa}  

\usepackage{graphicx}
\usepackage{txfonts}
\usepackage{hyperref}
\begin{document}

   \title{20 years of photometric microlensing events\\predicted by \emph{Gaia} DR2}

   \subtitle{Potential planet-hosting lenses within 100 pc}

   \author{Alexander J. Mustill
          \inst{1}
          \and
          Melvyn B. Davies\inst{1}
          \and
          Lennart Lindegren\inst{1}
          }

   \institute{$^1$Lund Observatory, Department of Astronomy \& Theoretical Physics,
     Lund University, Box 43, SE-221 00 Lund, Sweden\\
     \email{alex@astro.lu.se}
   }

   \date{Received 29 May 2018; accepted XXX}

 
  \abstract
   {\emph{Gaia} DR2 offers unparalleled precision on stars' parallaxes 
     and proper motions. This allows the prediction of microlensing events 
     for which the lens stars (and any planets 
     they possess) are nearby and may be well 
     studied and characterised.}
   {We identify a number of potential microlensing events that will 
   occur before the year 2035.5, 20 years from the \emph{Gaia} DR2 reference epoch.}
   {We query \emph{Gaia} DR2 for potential lenses within 
   100~pc, extract parallaxes and proper motions of the lenses 
   and background sources, and identify potential lensing events. 
   We estimate the lens masses from Priam effective temperatures, 
   and use these to calculate peak magnifications and the 
   size of the Einstein radii relative to the lens stars' 
   habitable zones.}
   {We identify 7 future events with a probability $>10\%$ of an alignment 
     within one Einstein radius. Of particular interest 
     is DR2~5918299904067162240 (WISE J175839.20–583931.6), 
     magnitude $G=14.9$, which will lens 
     a $G=13.9$ background star in early 2030, with a median 23\% net magnification.
     Other pairs are typically fainter, 
     hampering characterisation of the lens (if the lens is faint) or the 
     ability to accurately measure the magnification (if the source is 
     much fainter than the lens). Of timely interest is 
     DR2~4116504399886241792 (2MASS J17392440–2327071), 
     which will lens a background star in 
     July 2020, albeit with weak net magnification ($0.03\%$). 
     Median magnifications for the other 5 
     high-probability events range from $0.3\%$ to $5.3\%$. 
     The Einstein radii for these lenses are 1--10 times the 
     radius of the habitable zone, allowing these lensing events 
     to pick out cold planets around the ice line, and 
     filling a gap between transit and current microlensing 
     detections of planets around very low-mass stars.}
   {We provide a catalogue of the predicted events to aid future 
     characterisation efforts. Current limitations include a lack of 
     many high-proper motion objects in \emph{Gaia}~DR2 and often large 
     uncertainties on the proper motions of the background 
     sources (or only 2-parameter solutions). Both of these deficiencies 
     will be rectified with \emph{Gaia}~DR3 in 2020. Further characterisation 
     of the lenses is also warranted to better constrain their masses and 
     predict the photometric magnifications.}


   \maketitle
%

\section{Introduction}

Gravitational microlensing 
\citep{Einstein36,Paczynski86a,Paczynski96} 
allows the detection of low-mass planets 
on orbits in or beyond a star's habitable zone 
\citep{Bond+04,Gaudi12,Shvartzvald+16,Suzuki+16}. 
However, current microlensing 
searches are hampered by small probability of alignments 
and the need to survey many stars, which are typically 
distant and therefore faint. 
Indeed the lens star may not be known at all, if it is 
an M dwarf at a few kpc. Characterisation of the 
host and the planetary system is thus often probabilistic 
conditioned over a model of Galactic stellar populations 
\citep[see, e.g., ][for recent examples]{Hwang+18,Jung+18a}.

While detections of microlensing by planet-hosting stars 
have so far been conducted by wide-field surveys searching 
for chance alignments (such as OGLE, \citealt{Udalski03}; 
MOA, \citealt{Bond+01}; and KMTNet, \citealt{Kim+16}), the alignments are themselves 
determined by the proper motion and parallax of both the 
source and the lens. With sufficiently accurate 
astrometry, future lensing events can therefore be predicted 
in advance, allowing both observations of the lensing event to be 
optimised, and efforts made to characterise the lensing star 
and any planetary system \citep{Paczynski95,DiStefano08,DiStefano+13}. 
The necessary astrometry for predicting future lensing events is provided by
ESA's ongoing \emph{Gaia} mission \citep{GaiaMission16}, which
provides proper motions and parallaxes to
excellent precision for stars covering the entire sky.
In this paper, we show that the second data release
\citep[``DR2'',][]{GaiaSummary18} does indeed allow the prediction of future
microlensing events, and provide a list of candidates and
predicted magnifications.
Previous work has used pre-\emph{Gaia} 
or DR1 astrometry to predict lensing events by the 
nearby $\alpha$~Cen system \citep{Kervella+16} or by known stellar 
remnants \citep{Paczynski01,Harding+18,McGill+18}, often 
focusing on the astrometric signature. However, before 
\emph{Gaia} DR2 the proper motion of the background sources in 
particular was unknown, hampering accurate predictions 
of the microlensing signal. 
Recently, \cite{Klueter+18} identified two ongoing astrometric 
events from \emph{Gaia}~DR2. In this paper, we focus 
on photometric signatures from lensing events 
up to 2035 that can be predicted 
(with associated uncertainties) from \emph{Gaia} DR2\footnote{On the day of submission of
this paper we became aware of a complementary study by \cite{Bramich18},
which covers events up to mid 2026.}.

Nearby lens stars are of particular interest, but are 
not well represented in wide-field surveys. Because 
of their higher proper motions, they cover a larger 
fraction of sky and have the opportunity to lens more 
background sources. Because of their proximity, they 
are bright and therefore easier to characterise, or 
to be probed for planets by complementary methods. 
Ideally the background source star is not too faint however, 
as a large magnitude difference between source and lens will 
dilute the photometric microlensing signal and make 
characterisation of the lens system (including any planets) 
through the microlensing light curve more challenging.

In this paper we search for alignments between sources and lenses 
in \emph{Gaia}~DR2 \citep{GaiaSummary18}, and calculate the 
magnitude of the photometric signal for each pair. We focus on lenses 
within 100pc. We describe our sample selection and analysis 
method in Section~\ref{sec:method}, present results 
in Section~\ref{sec:results}, and discuss and conclude 
in Section~\ref{sec:discuss}. First we briefly review 
the theoretical background of gravitational microlensing.

A brief note on terminology: the \emph{Gaia} data releases
refer to ``sources''. However, in microlensing it is
convenient to refer to a foreground ``lens'' and a background
``source''. Here we use the microlensing terminology,
and refer to ``sources'' in \emph{Gaia}~DR2 as ``objects''.

\section{Microlensing theory}

If a foreground lens star lies exactly on a line linking 
an observer and a background source, the light of the source 
is distorted into a ring of angular radius 
\begin{eqnarray}
  \theta_\mathrm{E} & = &\left(\frac{4\mathcal{G}M}{D_\mathrm{rel}c^2}\right)^{1/2}\\
  \frac{\theta_\mathrm{E}}{\mathrm{mas}} & = &2.9\left(\frac{M}{\mathrm{M}_\odot}\right)^{1/2}
                         \left(\frac{\varpi_\mathrm{L}}{\mathrm{mas}}-\frac{\varpi_\mathrm{S}}{\mathrm{mas}}\right)^{1/2},
\end{eqnarray}
the angular Einstein radius \citep{Einstein36,Paczynski86a,Paczynski96}. 
Here, $\mathcal{G}$ is 
the gravitational constant, $M$ the lens mass, $c$ the speed of 
light, and $D_\mathrm{rel}^{-1}=D_\mathrm{L}^{-1}-D_\mathrm{S}^{-1}$, 
where $D_\mathrm{L}$ and $D_\mathrm{S}$ are the distances to the source and the lens, 
and $\varpi_\mathrm{L}$ and $\varpi_\mathrm{S}$ their parallaxes. 
This angular radius corresponds to a physical Einstein radius 
\begin{eqnarray}
  R_\mathrm{E} & = & \theta_\mathrm{E}D_\mathrm{L}\\
  \frac{R_\mathrm{E}}{\mathrm{au}} & = & 2.9\left(\frac{M}{\mathrm{M}_\odot}\right)^{1/2}
   \left(\frac{\varpi_\mathrm{L}}{\mathrm{mas}}-\frac{\varpi_\mathrm{S}}{\mathrm{mas}}\right)^{1/2}
   \left(\frac{\varpi_\mathrm{L}}{\mathrm{mas}}\right)^{-1},
\end{eqnarray}
or, when the lens is nearby (source parallax negligible)
\begin{equation}
  \frac{R_\mathrm{E}}{\mathrm{au}} = 0.29\left(\frac{M}{\mathrm{M}_\odot}\right)^{1/2}
  \left(\frac{\varpi_\mathrm{L}}{100\mathrm{\,mas}}\right)^{-1/2}.
\end{equation}
For a lens offset from the source--observer axis by 
$\theta_\mathrm{S}$, the 
light from the source is magnified by a factor 
\begin{equation}
  A = \frac{u^2+2}{u\sqrt{u^2+4}},
\end{equation}
where $u=\theta_\mathrm{S}/\theta_\mathrm{E}$. 
A separation of one Einstein radius thus results 
in a magnification of $3/\sqrt{5}\approx1.34$. 

For many microlensing applications the source is a 
background giant and much brighter than the lens, 
which is typically an M dwarf at a few kpc. Dilution 
of the source light by the lens is therefore usually 
small. As we shall see, however, the source--lens pairs 
studied here have lenses comparable to or brighter than 
the background sources, and when discussing 
the practical observability of the events we 
study we need to calculate the net magnification 
including dilution by the lens itself. We define 
\begin{equation}
  A_\mathrm{net} = \frac{1+AF_\mathrm{S}}{1+F_\mathrm{S}}-1
\end{equation}
where
\begin{equation}
  F_\mathrm{S} = 10^{0.4\left(G_\mathrm{L}-G_\mathrm{S}\right)},
\end{equation}
$G_\mathrm{L}$ and $G_\mathrm{S}$ being 
the unmagnified  $G$-band magnitudes 
of the lens and the source.

The timescale of a microlensing event is usually 
parametrized by the Einstein ring crossing 
time
\begin{equation}
  t_\mathrm{E} = \theta_\mathrm{E}/\mu_\mathrm{rel}
\end{equation}
\citep{Gaudi12}, 
where $\mu_\mathrm{rel}$ is the relative proper motion 
of the source and the lens. Because all of our events are 
significantly diluted by the lens light, however, in 
this paper we instead parametrize the duration 
by the length of time the net magnification 
exceeds $10^{-4}$, a reasonable estimate for 
the precision attainable by future high-precision photometry 
such as that from \emph{PLATO} (see Section~\ref{sec:discuss}).

\section{Method}

\label{sec:method}

\subsection{Sample selection}

We initially queried for objects within 100pc of the Sun 
(nominal parallax $\varpi\ge10$\,mas) from the \emph{Gaia} DR2 
archive. This returned 700\,055 objects. We then applied the 
criteria on photometric and astrometric quality described 
in \cite{Lindegren+18}, Appendix~C. This reduced the list to 
69\,979 objects. For each of these objects, we then queried for 
all sources within a radius of 
$20\sqrt{\mu_{\alpha*}^2+\mu_{\delta}^2} + 2\varpi$, 
where $\mu_{\alpha*}$ and $\mu_{\delta}$ are the object's annual proper 
motion components in RA and Dec and $\varpi$ the parallax. That is, 
we search for objects within a radius given by 20 years' proper motion
plus twice the parallax. We find that 2631 objects have at 
least one other object in this search radius.

\begin{figure}
  \includegraphics[width=0.5\textwidth]{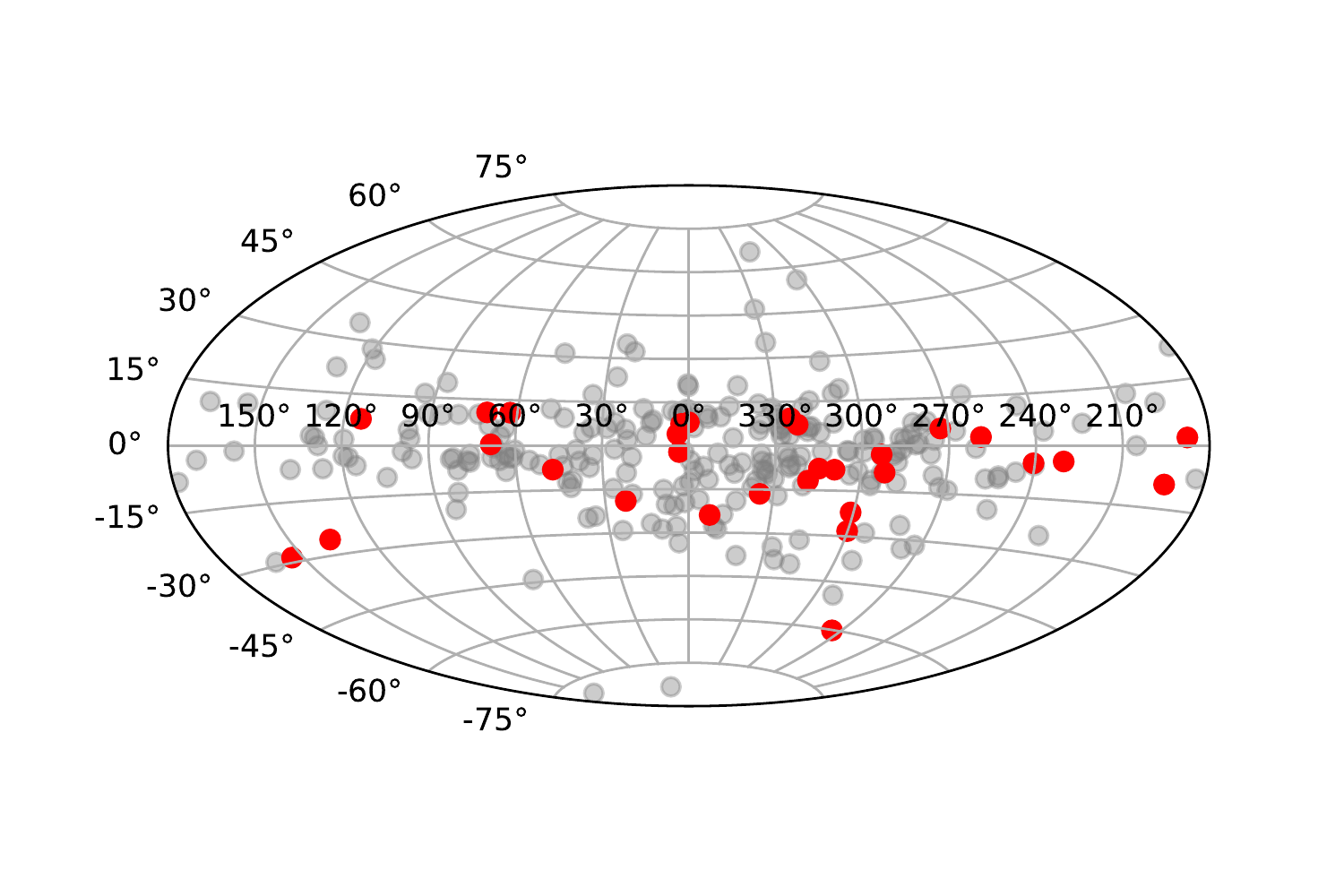}
  \caption{Sky locations (Galactic co-ordinates, 
    Aitoff projection) of the 
    262 objects within 100\,pc passing within 
    1 arcsec of a background source. A 
    concentration towards the Galactic plane 
    is noted. The 30 objects passing 
    within 0.1 arcsec of a background source 
    are highlighted in red.}
  \label{fig:aitoff}
\end{figure}

From these, we select objects whose nominal proper motion and parallax takes them 
within 1 arcsec of another object for further analysis\footnote{Earth 
ephemeris data for the parallax are taken from the JPL Horizons system 
\url{https://ssd.jpl.nasa.gov/?horizons} \citep{Giorgini+96}.}. This 
gives 262 objects which are then propagated with the nominal parallax and 
proper motion of themselves and the background objects. Objects passing within 
0.1 arcsec of another object are then selected for further manual study. 
There were 30 such objects, listed Table~\ref{tab:all} 
in the Appendix. The sky locations of these 
objects are shown in Figure~\ref{fig:aitoff}.

\subsection{Analysis}

For the 30 objects passing within 0.1 arcsec of a background 
source, we perform a Monte-Carlo simulation of 10\,000 samples to estimate 
the distribution of closest approaches. We use the formal 
errors on parallax, proper motion, RA and Dec given 
in the DR2 catalogue, and account for the correlations between 
these parameters \citep{Luri+18}. Background sources for which only two 
astrometric parameters are available (RA and Dec) have 
these parameters randomised but are assigned zero 
parallax. Proper motions are then assigned by 
querying for stars within 1 magnitude of the background source in a 
1 arcmin radius, and drawing from Gaussians with the 
mean and standard deviation of this sample (typically $\sim200$ objects).
If a negative parallax is drawn for the 5-parameter 
objects, it is set to zero. 
For our purposes here, possible global 
or local systematics are unimportant, as they take effect on 
much larger scales than those of a few arcsec considered in 
this paper; in fact, \emph{relative} proper motions 
(considered here) may be slightly more accurate 
for very close pairs than the formal errors suggest.

Our Monte-Carlo analysis allows us to derive probability 
distributions for the trajectories on the sky of the 
source and the lens, the distribution of Einstein radii, 
of closest approaches, 
of the photometric magnifications by lensing, and 
the probability that a closest approach is within 
$1R_\mathrm{E}$. To compute the Einstein radius, 
we used the $T_\mathrm{eff}$ provided by the Priam 
algorithm \citep[part of DR2,][]{Andrae+18} where available, which 
together with the extinction and bolometric 
correction (Table 4 of \citealt{Andrae+18}) 
we turned into a bolometric 
luminosity $L$, and thence into a mass by 
the mass--luminosity relations of 
\cite{SalarisCassisi05}. Where $T_\mathrm{eff}$ 
is unavailable, we assume a lens mass of $0.1\mathrm{\,M}_\odot$. 
Our lens masses calculated in this manner lie in the range 
$0.09-0.30\mathrm{\,M}_\odot$. However, the lens masses 
calculated in this manner are not monotonic if $T_\mathrm{eff}$, 
and we emphasize the need for follow-up photometry or spectroscopy 
to better determine these stars' masses and refine 
the predictions for Einstein radii and lensing magnifications.

The lens co-ordinates were then checked in SIMBAD and 
VizieR to find other identifiers (if any) for these objects.

\section{Results}

\label{sec:results}

\begin{figure}
  \includegraphics[width=0.5\textwidth]{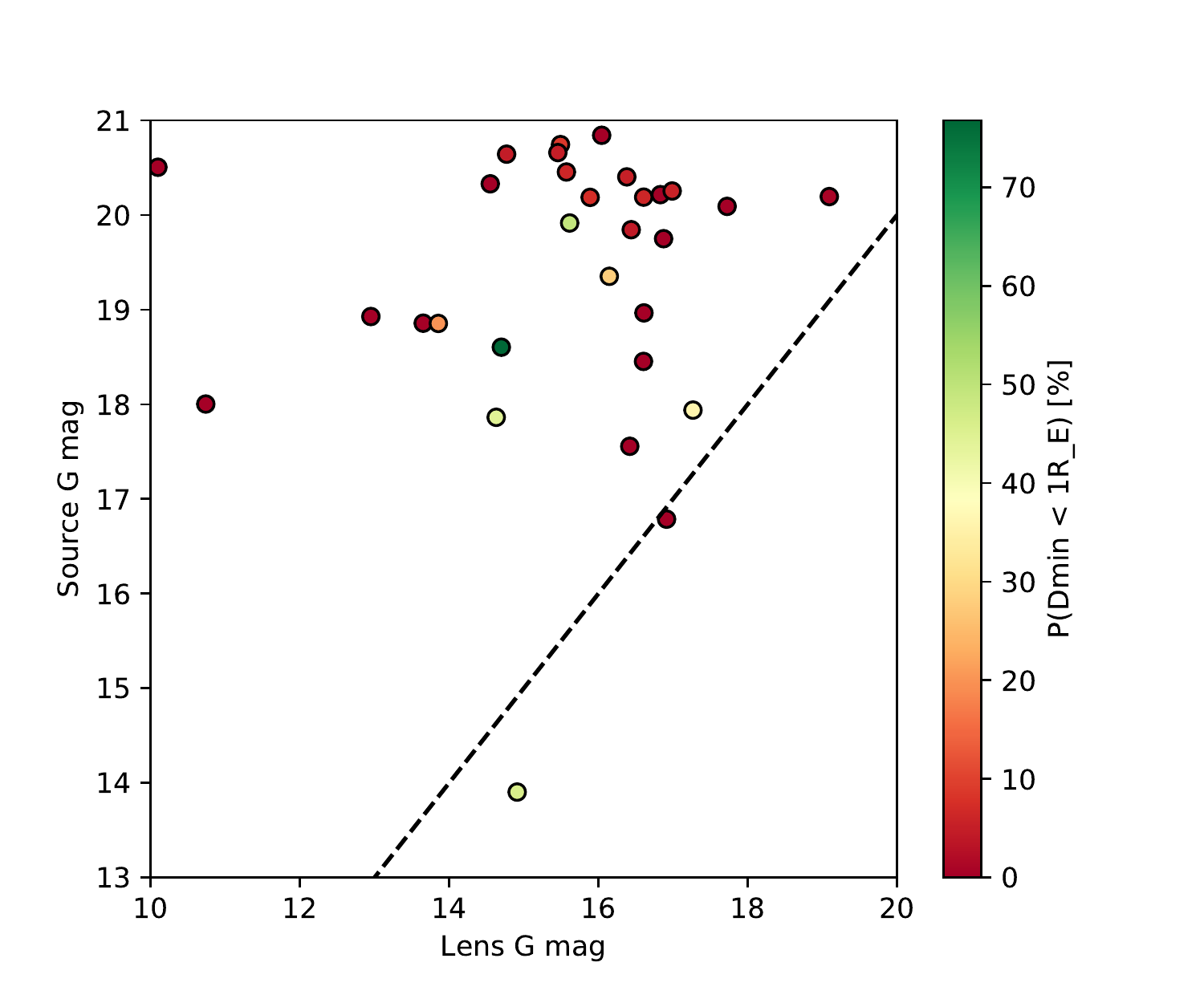}
  \caption{Magnitudes of sources and lenses of the 30 pairs 
    passing within 0.1 arcsec by AD2035.5. The colour scale
    shows the probability of an alignment within 1 Einstein 
    radius, taking the lens mass from Priam \citep{Andrae+18}. 
    The dashed line shows the locus of equal source and lens 
    magnitudes.}
  \label{fig:mag_mag_prob}
\end{figure}

\begin{figure*}
  \includegraphics[width=0.5\textwidth]{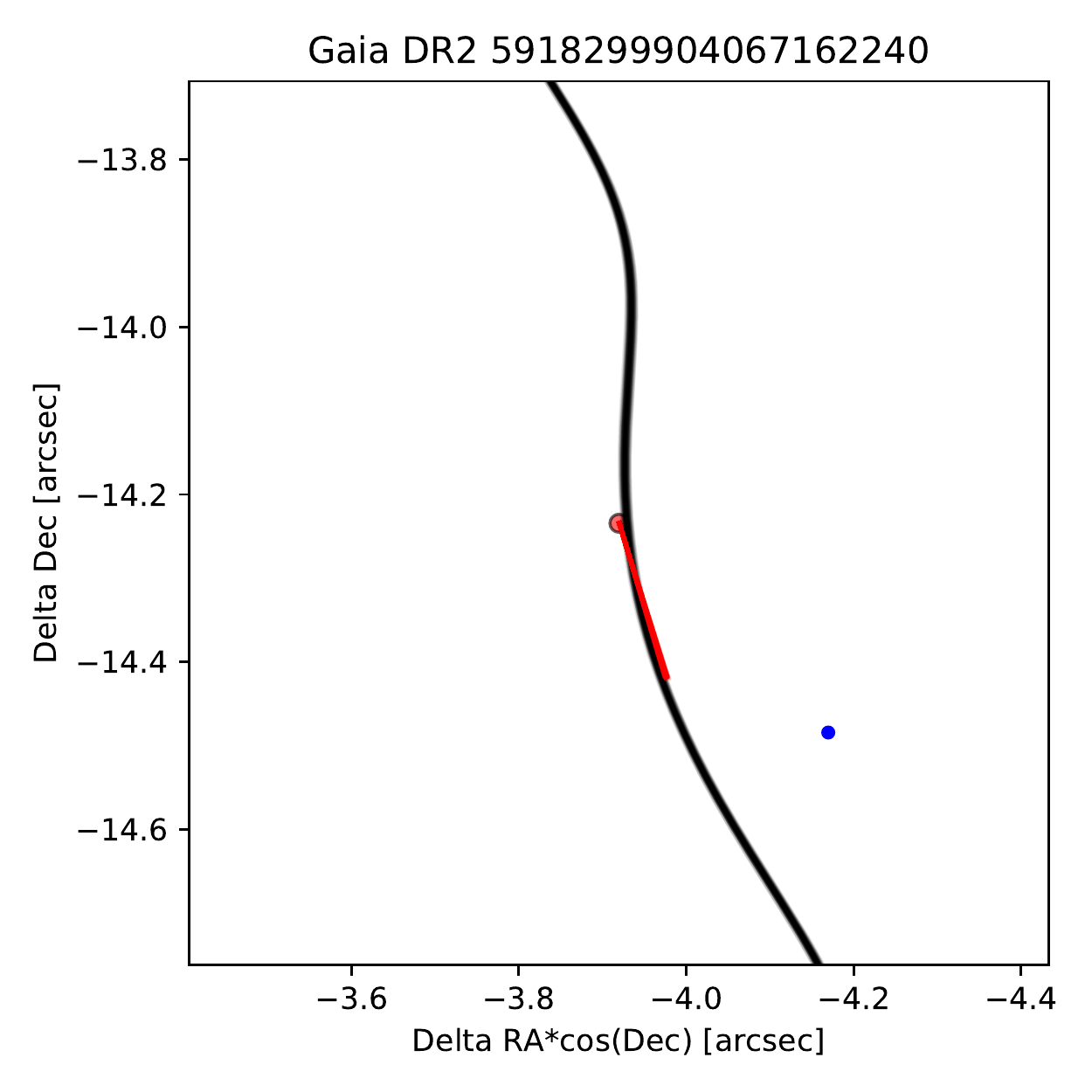}
  \includegraphics[width=0.5\textwidth]{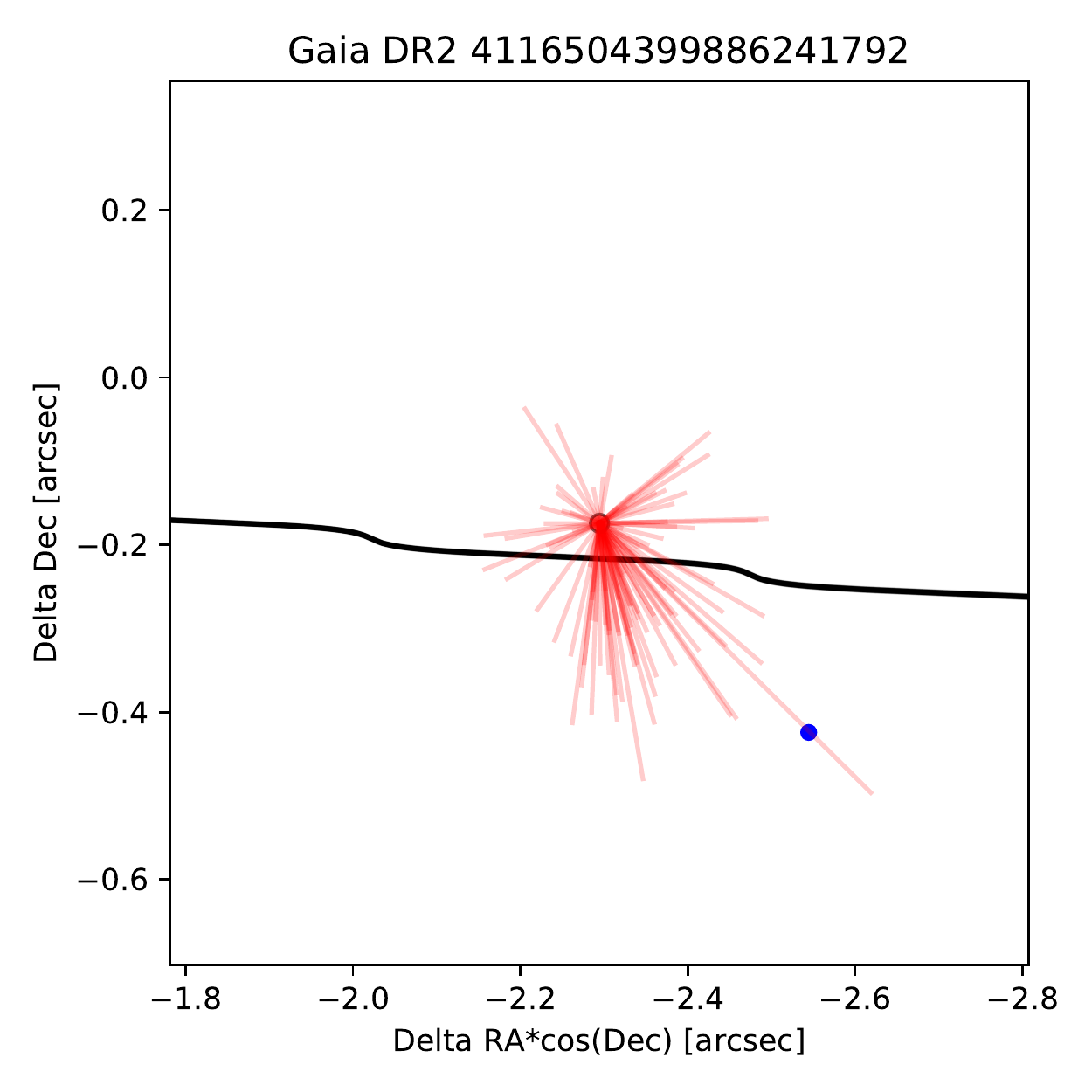}
  \caption{Sky motion of two candidate events. Lens motion is marked in 
    black (moving southwards and westwards in each case) and source motion 
    from 2015.5 to 2035.5 in red 
    (moving away from the red circle, showing the source position at 
    2015.5), for 100 draws from the Monte-Carlo samples. RA and Dec 
    are given relative to the lens location at 2015.5. 
    Left: DR2~5918299904067162240 (lens) and DR2~5918299908365843840 
    (source). Both source and lens motion are well-constrained, and 
    there is a 46\% probability of passing within $1\mathrm{\,R_E}$, 
    around 1 April 2030. 
    Right: DR2~4116504399886241792 (lens) and DR2~4116504399831319424 
    (source). The background source has only a 2-parameter 
    astrometric solution; 
    by assigning it a proper motion drawn from stars in 
    the same field of similar magnitude, we can 
    predict a 20\% probability of passing within 
    $1\mathrm{\,R_E}$, around July 2020.
    In each panel, the Einstein radius is indicated by 
    a blue circle.}
  \label{fig:radec}
\end{figure*}

We identified 30 sources in \emph{Gaia}~DR2 which will 
pass within 0.1~arcsec of a background object before 
AD2035.5. These alignments are concentrated towards 
the Galactic plane where the density of background sources 
is higher (Figure~\ref{fig:aitoff}), but a few alignments occur 
at high Galactic latitudes. 

Lenses are typically brighter than the background sources, as 
is expected from their closer proximity to Earth. All but 
two background sources are fainter than $G=17$, while all 
but three lenses are brighter than $G=17$. For two alignments 
the background source is brighter than the lens; one of these 
(DR2~5918299908365843840) we discuss in more detail below. The magnitudes of the 
source--lens pairs are plotted in Figure~\ref{fig:mag_mag_prob}, 
along with the probability of an alignment within 
$1R_\mathrm{E}$.

\begin{table*}
  \caption[]{The 7 most probable alignments before 2035.5, ordered chronologically. We give \emph{Gaia}~DR2 
    identifiers for the lens and the source, the probability of passage within 
    1 Einstein radius $P(d<1R_\mathrm{E})$, the median and 68\% confidence intervals 
    for the peak net magnification (including dilution by the lens), and the epoch of peak magnification. 
    Alternative identifiers for the lenses are given in the footnotes (marked `?' where not certain). 
    Further details for these and all 30 
    pairs are given in the Appendix.}
  \label{tab:best}
  
  \begin{tabular}{ccccccc}
    \hline
    \noalign{\smallskip}
    Lens DR2 id & Lens $G$ & Source DR2 id & Source $G$ & $P(d<1R_\mathrm{E})$ & $f_\mathrm{max}$ & $T_\mathrm{max}$ \\
    \noalign{\smallskip}
    \hline
    \noalign{\smallskip}
    4116504399886241792\tablefootmark{a} & 13.86 & 4116504399831319424\tablefootmark{b} & 18.85 & 19.95 & $2.86^{+49.70}_{-2.56}\times10^{-4}$ & 2020 Jul 3--22\\
    2042703905028908928\tablefootmark{c} & 14.70 & 2042703900727995008 & 18.60 & 76.77 & $2.13^{+7.59}_{-1.43}\times10^{-2}$ & 2025 Jul 20--27 \\
    2030898139472914688 & 15.62 & 2030898105088410112 & 19.92 & 49.29 & $6.19^{+36.20}_{-5.23}\times10^{-3}$ & 2025 Sep 1--16 \\
    5903487940560263936\tablefootmark{d} & 14.63 & 5903487940560263808 & 17.86 & 44.60 & $1.34^{+6.31}_{-0.96}\times10^{-2}$ & 2026 May 25 -- Jun 7 \\
    318399257231042304\tablefootmark{e,f} & 17.27 & 318399257231042048 & 17.94 & 35.85 & $5.31^{+45.34}_{-4.58}\times10^{-2}$ & 2028 Jul 10--27 \\
    5918299904067162240\tablefootmark{g} & 14.91 & 5918299908365843840 & 13.90 & 45.93 & $2.27^{+2.76}_{-1.11}\times10^{-1}$ & 2030 Mar 31 -- Apr 2 \\
    3425318817510655360\tablefootmark{h} & 16.15 & 3425318813215052288 & 19.35 & 27.96 & $3.42^{+42.39}_{-3.05}\times10^{-3}$ & 2034 Sep 7--19 \\
    \noalign{\smallskip}
    \hline
  \end{tabular}
\tablefoottext{a}{2MASS J17392440--2327071.}
\tablefoottext{b}{2-parameter astrometric solution only.}
\tablefoottext{c}{G 207--14}.
\tablefoottext{d}{USNO-B1.0 0421--0506943?}
\tablefoottext{e}{LSPM J0146+3545.}
\tablefoottext{f}{No $T_\mathrm{eff}$ from Priam.}
\tablefoottext{g}{WISE J175839.20--583931.6.}
\tablefoottext{h}{2MASS J06095230+2319143.}
\end{table*}

We found 18 such alignments with a nonzero probability 
of passing within $1R_\mathrm{E}$, including seven with 
a probability $>10\%$. Details for 
these seven are given in Table~\ref{tab:best} and for all 30 
pairs in the Appendix. Here we describe the seven likely lensing 
events, in chronological order.

\subsection{2020 Jul 3--22: DR2~4116504399886241792}

The first potential lensing event will occur two years from now. 
\emph{Gaia} DR2~4116504399886241792 will pass close to 
the background object \emph{Gaia} DR2~4116504399831319424 in 
mid 2020 (Figure~\ref{fig:radec}, right). Unfortunately the background object has only 
a 2-parameter astrometric solution in DR~2; by assigning 
it a proper motion based on stars of similar magnitude in the field 
we predict a 20\% probability of an alignment within 1 Einstein 
radius, with a peak net magnification of $2.86^{+49.70}_{-2.56}\times10^{-4}$ occuring 
around 2020 Jul 3--22. At this time, the Solar elongation will 
be $155^{+9}_{-10}$ degrees, very favourable for observations. 
As this is before the release of \emph{Gaia} 
DR~3, further monitoring of the source to better constrain its proper 
motion is desirable. The lens itself is one of the brighter among 
our full set ($G=13.86$) and the brightest of the seven highest-probability 
lenses, aiding characterisation efforts. Based on position and 
proper motion, we identify the lens with 2MASS J17392440--2327071 
in the high-proper motion catalogue of \citep{Lepine08}, 
who classed it as a dwarf star. Priam gives 
$T_\mathrm{eff}=3554$\,K, whence we derive a mass of $0.16\mathrm{\,M}_\odot$. 

\subsection{2025 Jul 20--27: DR2~2042703905028908928}

\emph{Gaia} DR2~2042703905028908928 ($G=14.7$) will
pass close to \emph{Gaia} DR2~2042703900727995008 ($G=18.6$) in
late July 2025, and it has the highest probability of an 
approach within one Einstein radius (77\%). 
Despite the lens being 4 magnitudes brighter than the 
source, we predict a large peak net magnification of 
$2.13^{+7.59}_{-1.43}\times10^{-2}$. The Solar elongation at 
peak magnification will be 126 degrees. 
The lens can be identified with G~207--14 based on its position and
proper motion \citep{LepineShara05}. The Priam $T_\mathrm{eff}=3655$\,K,
and we estimate a lens mass of $0.17\mathrm{\,M}_\odot$.

\subsection{2025 Sep 1--16: DR2~2030898139472914688}

\emph{Gaia} DR2~2030898139472914688 ($G=15.6$) has a 49\% 
probability of passing within an Einstein radius of 
\emph{Gaia} DR2~2030898105088410112 ($G=19.9$), 
with a peak magnification of $6.19^{+36.20}_{-5.23}\times10^{-3}$ 
occurring in early September 2025, with a Solar 
elongation of $122^{+3}_{-3}$ degrees. Priam 
gives $T_\mathrm{eff}=4002$\,K and we derive a lens mass of 
$0.14\mathrm{\,M}_\odot$. 

\subsection{2026 May 25 -- Jun 7: DR2~5903487940560263936}

\emph{Gaia} DR2~5903487940560263936 ($G=14.6$) will pass close to
\emph{Gaia} DR2~5903487940560263808 ($G=17.9$) in late May 
or early June 2026, with a 45\% probability of passing 
within one Einstein radius. We predict a peak net magnification 
of $1.34^{+6.31}_{-0.96}\times10^{-2}$, and a favourable 
Solar elongation of $146^{+2}_{-3}$ degrees. Priam 
gives $T_\mathrm{eff}=4065$\,K for the lens, and 
we derive a mass of $0.14\mathrm{\,M}_\odot$. 
Based on position, proper motion and magnitude, 
the lens is likely USNO-B1.0~0421-0506943.

\subsection{2028 Jul 10--27: DR2~318399257231042304}

\emph{Gaia} DR2~318399257231042304 ($G=17.3$) will pass close to \emph{Gaia}
DR2~318399257231042048 ($G=17.9$) in mid--late July 2028, with a 
36\% probability of passing within one Einstein radius. 
This pair has a favourable
contrast ratio, and our second-highest peak net 
magnification of $5.31^{+45.34}_{-4.58}\times10^{-2}$. 
However, the lens itself is faint, which will hamper
characterisation of any planetary system, and the Solar 
elongation at peak magnification is only $79^{+8}_{-7}$ degrees. 
The lens has no Priam solution for $T_\mathrm{eff}$, but has a red colour
($G_\mathrm{BP}-G_\mathrm{RP}=3.7$). Based on position and proper motion
the lens can be identified with LSPM~J0146+3545 \citep{LepineShara05}.

\subsection{2030 Mar 31 -- Apr 1: DR2~5918299904067162240}

\begin{figure}
  \includegraphics[width=0.5\textwidth]{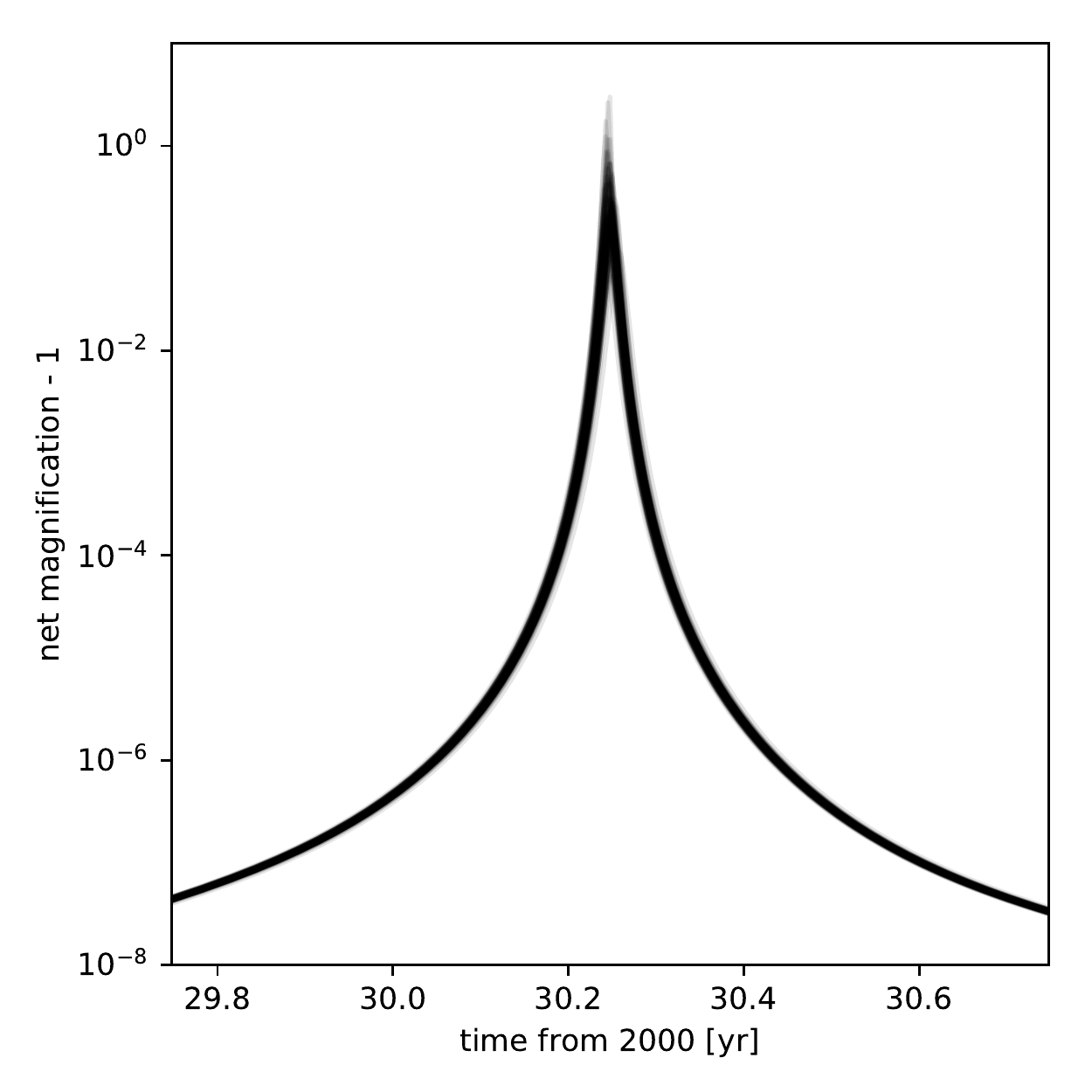}
  \caption{Net magnification of the background source 
  DR2~5918299908365843840 by the lens 
  DR2~5918299904067162240 in early 2030. Magnification 
  (including dilution by the lens) is shown for 
  100 Monte-Carlo samples, for 1 yr around the peak.}
  \label{fig:magnification}
\end{figure}

\emph{Gaia} DR2~5918299904067162240 is a $G=14.9$ star 
which will make its closest approach to the $G=13.9$ background source 
DR2~5918299908365843840 on or close to 1 April 2030. 
We find a 46\% probability of passing within one 
Einstein radius, and thanks to the background source being 
1 magnitude brighter than the lens we find a 
large peak net magnification of 
$2.27^{+2.76}_{-1.11}\times10^{-1}$. 
Solar elongation at peak magnification will be 99 degrees. 
Priam gives $T_\mathrm{eff}=3613$\,K for the lens, 
and we derive a mass of $0.1\mathrm{\,M}_\odot$. 
The lens can be identified with
\object{WISE J175839.20-583931.6}, classed as
a mid-M dwarf by \cite{LuhmanSheppard14}, so 
the mass (and therefore the 
peak magnification) might be higher.
The source has $T_\mathrm{eff}=4964$\,K and a small parallax
of $0.22\pm0.03$\,mas.

\subsection{2034 Sep 7--19: DR2~3425318817510655360}

The last of our high-probability alignments, 
\emph{Gaia} DR2~3425318817510655360 ($G=16.1$) will pass close to 
\emph{Gaia} DR2~3425318813215052288 ($G=19.4$) in mid September 2034, 
with a 28\% probability of coming within 1 Einstein radius. 
The lens is fairly faint, and the net magnification also 
fairly low at $3.42^{+42.39}_{-3.05}\times10^{-3}$, 
while the Solar elongation will be only 
$78^{+6}_{-6}$ degrees. The lens may be identified
with LSPM~J0609+2319 based on position and proper motion,
an M5.0 subdwarf \citep{Lepine+03}. From the Priam 
$T_\mathrm{eff}=3546\,K$ we derived a lens mass of 
$0.15\mathrm{\,M}_\odot$.

\section{Discussion}

\label{sec:discuss}

\begin{figure}
  \includegraphics[width=0.5\textwidth]{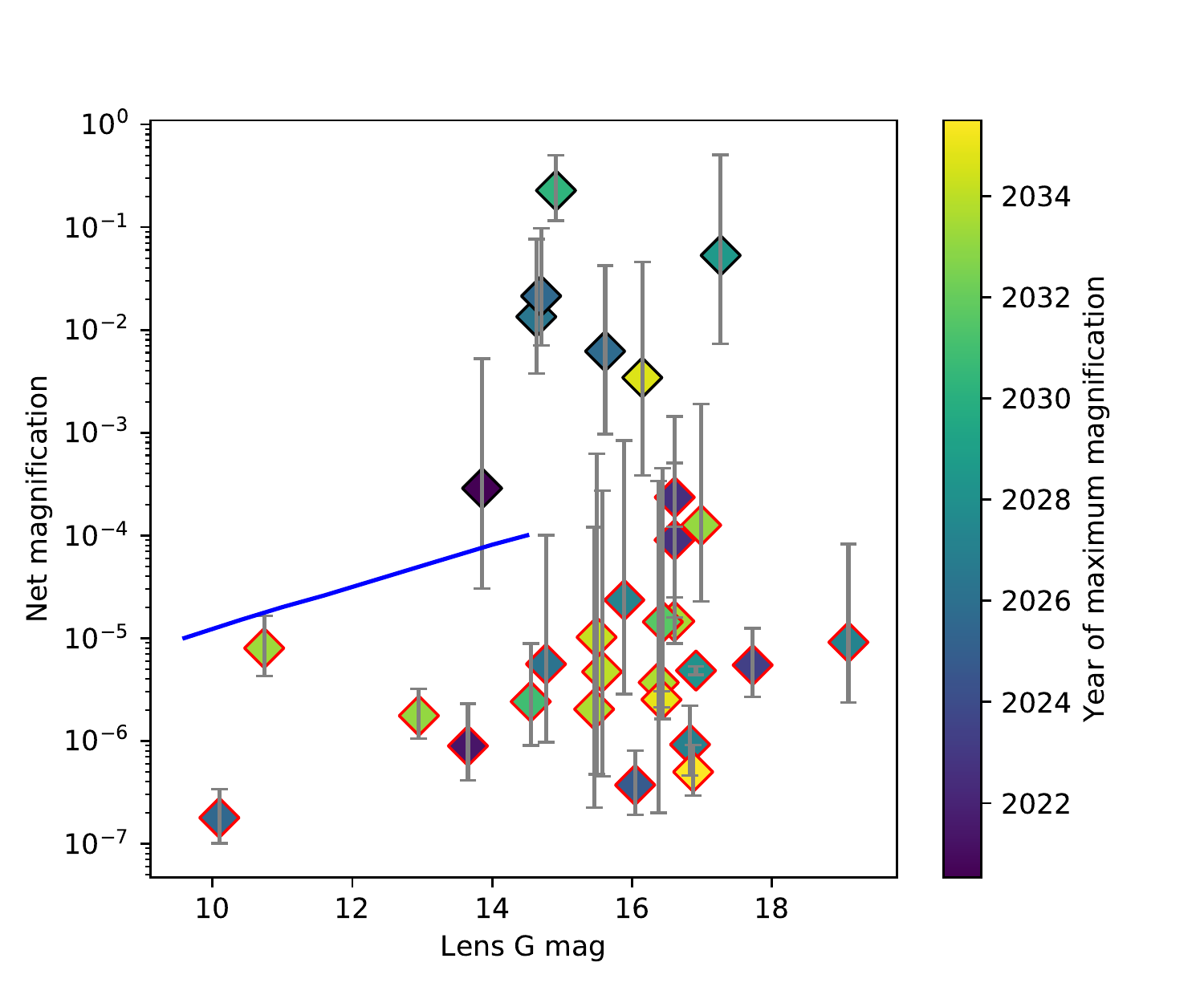}
  \caption{$G$ magnitude of lens and net magnification, 
    accounting for dilution by the lens. 
    Fill colour of markers shows the epoch of 
    maximum magnification. Markers outlined in black 
    mark the high-probability events listed in 
    Table~\ref{tab:best}; the remainder are outlined in red. 
    The blue line shows approximate 
    \emph{PLATO} precision in 1\,hr of observations \citep{Catala09}.}
  \label{fig:Gmag_magnification}
\end{figure}

From \emph{Gaia} DR2 we have identified 
30 approaches of lenses within 0.1 arcsec
of background sources before 2035.5. 
Seven lenses have a $>10\%$ probability 
of approaching within $1R\mathrm{E}$ of the background 
source. The lens $G$-band magnitudes vary from 
10.1 to 19.1, with the seven most probable 
in the range 13.9 to 17.3. Typically the source is 
considerably fainter than the lens (Figure~\ref{fig:mag_mag_prob}), 
but nevertheless significant net magnifications of 
up to 20\% are possible for the high-probability 
pairs. The lens $G$ magnitude and the peak net magnification 
are plotted in Figure~\ref{fig:Gmag_magnification}, 
where we also show the nominal photometric 
precision attainable by \emph{PLATO} \citep{Catala09,Rauer+14} 
as an example of future high-precision photometric performance.

\begin{figure}
  \includegraphics[width=0.5\textwidth]{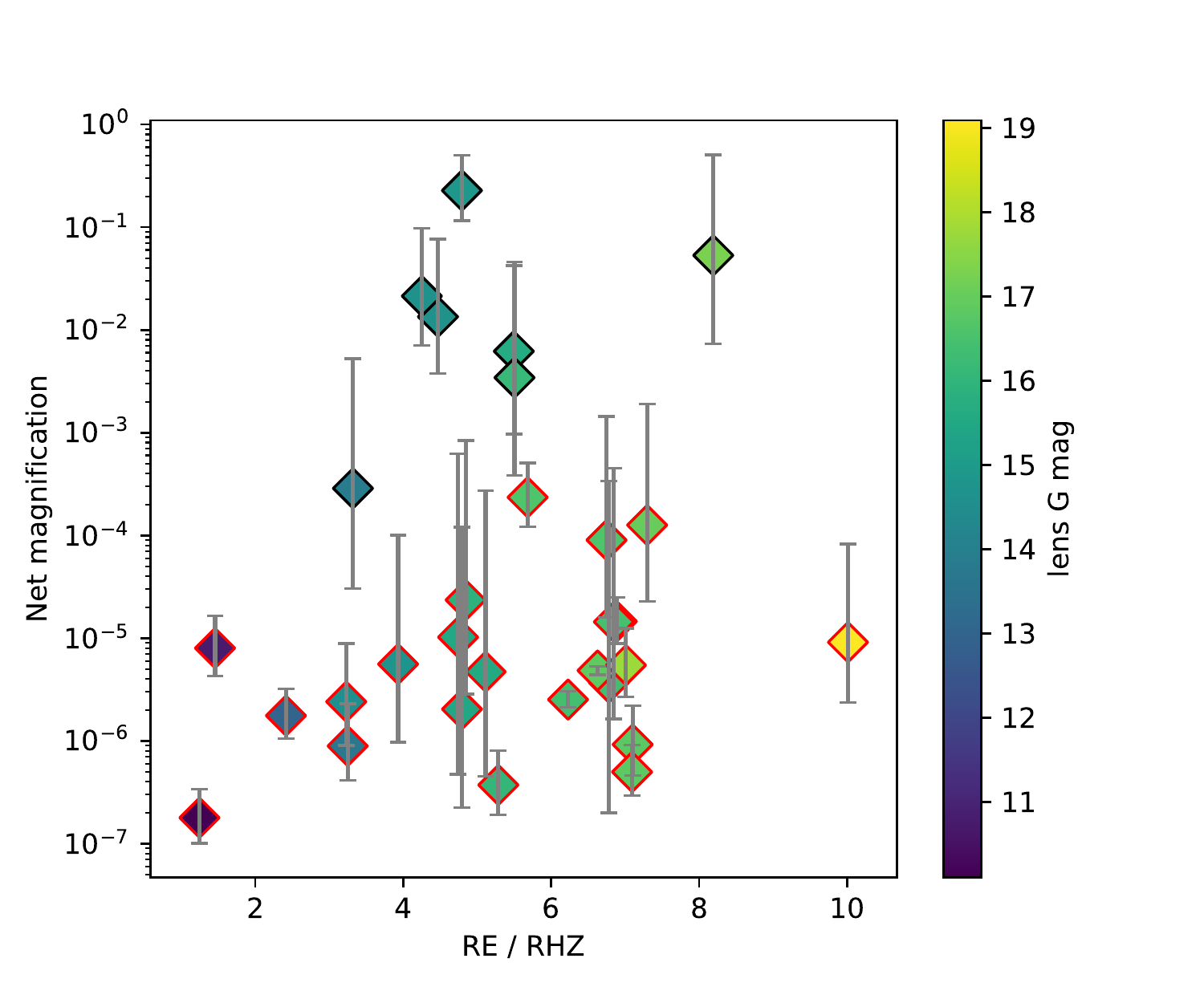}
  \caption{Net magnification of each event, plotted against 
    the Einstein radius scaled by the habitable zone width. 
    Fill colour shows the $G$ magnitude of the lens. 
    Marker outlines are as in 
    Figure~\ref{fig:Gmag_magnification}.}
  \label{fig:RE_netmag}
\end{figure}

Microlensing is most sensitive to planets 
located at $\sim1R_\mathrm{E}$ \citep{Gaudi12}. 
Because the lens stars are small and nearby, 
their Einstein radii are less than 1\,au 
in size. 
The estimated masses of the lens stars 
are in the range $0.09-0.3\mathrm{\,M}_\odot$, giving 
Einstein radii in the range $0.11-0.36$\,au. This is 
a factor 1--10 larger than the stars' habitable zones 
$R_\mathrm{HZ}$ (defined such that the insolation 
at that orbital radius is equal to that received 
by Earth). This distance is shown in 
Figure~\ref{fig:RE_netmag}, together with the 
peak net magnification. However, we note that 
further characterisation of the lens stars 
is necessary to refine their habitable zones, 
Einstein radii, and indeed the peak magnifications, 
since the errors on $T_\mathrm{eff}$ and extinction 
from Priam are somewhat large.

\begin{figure}
  \includegraphics[width=0.5\textwidth]{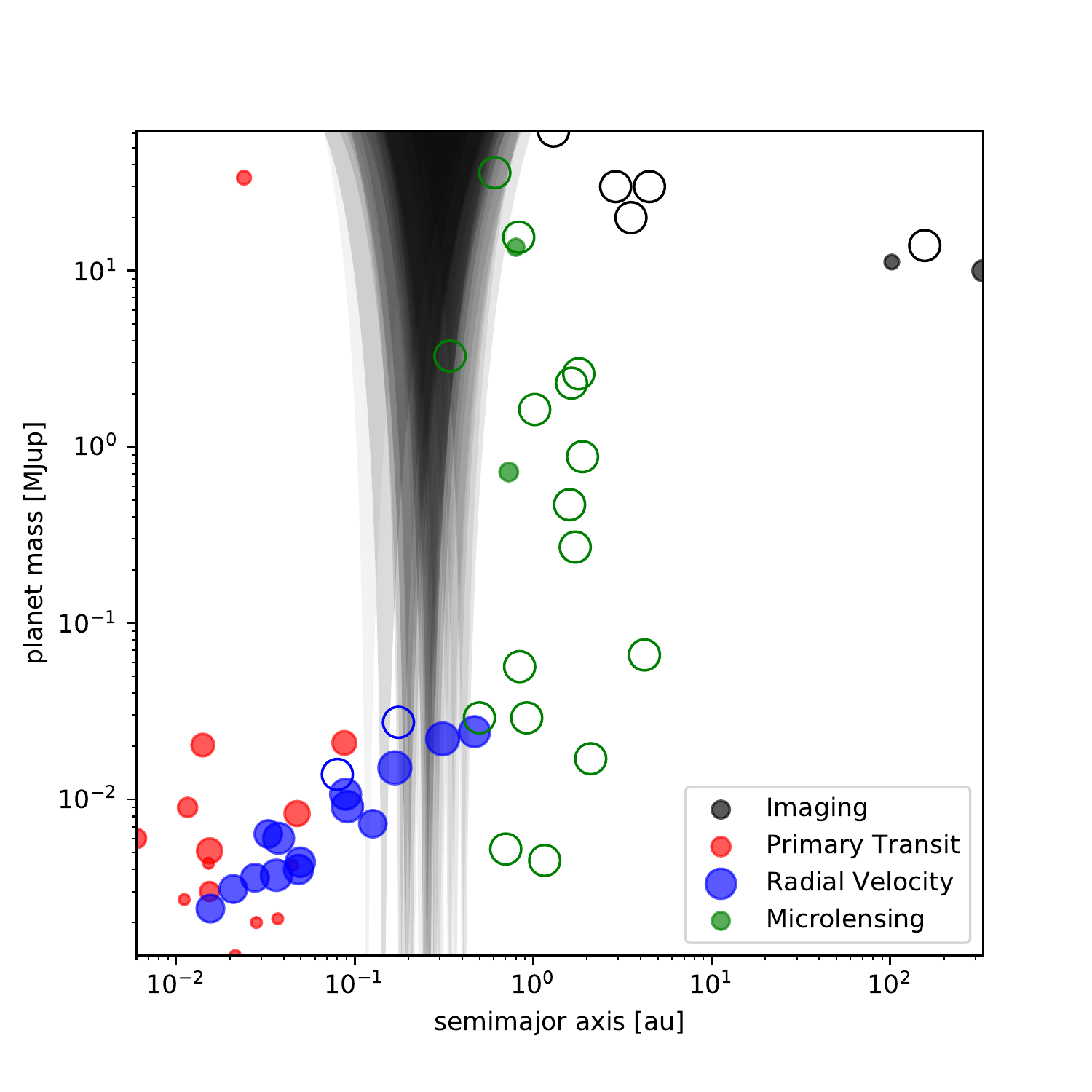}
  \caption{Currently-known planetary and brown dwarf 
      companions to low-mass ($0.07-0.3\mathrm{\,M}_\odot$) 
      stars; data from \url{http://exoplanet.eu/} on 
      2018~Jun~20. Symbol colour shows discovery method: 
      black for imaging, red for transit, blue for radial 
      velocity and green for microlensing. The symbol size 
      increases for lower $V$-band magnitudes; open 
      symbols lack a $V$-band magnitude. The grey wedges 
      show the region planets close to the Einstein 
      radius causing resonant microlensing distortions 
      (see Equation~\ref{eq:resonant}). The transparency 
      of the wedges increases as the closest approach increases.}
  \label{fig:newPlanets}
\end{figure}

In Figure~\ref{fig:newPlanets} we show 
potential new planetary discoveries from 
monitoring these lenses, in the parameter 
space of orbital semimajor axis and mass 
of \emph{known} planets orbiting very low-mass 
stars. We took all planets and brown dwarfs 
orbiting primaries with nominal masses 
$0.07-0.3\mathrm{\,M}_\odot$ from the 
Extrasolar Planets Encyclopaedia 
(\url{http://exoplanet.eu/}) on 2018~Jun~20. 
To give a sense of the region around the 
Einstein radius where planets may be detectable by 
microlensing, we overlay wedges showing the region where planets 
induce ``resonant'' microlensing perturbations, 
which has a width around the Einstein radius of 
\begin{equation}
  \Delta\theta = \frac{9}{4}\left(\frac{M_\mathrm{pl}}{M}\right)^{1/3}\theta,\label{eq:resonant}
\end{equation}
where $M_\mathrm{pl}$ is the planet mass and $M$ the lens mass 
\citep{Dominik99}. This region of detectability lies 
interior to the population of planets currently 
detected by microlensing around these low-mass stars, whose 
hosts are often very poorly characterised or only estimated 
based on Galactic models \citep[e.g.,][]{Jung+18b}. The region 
also lies outside the region where transiting planets 
have been detected so far, which extends to $0.1$\,au. It 
is unlikely that significant numbers of planets will be discovered 
by transit beyond $0.1$\,au around such low-mass stars: \emph{TESS}, 
for example, covers spectral types down to M5, missing late M dwarfs
\citep{Ricker+15}. The region of microlensing detectability 
overlaps with the most distant planets detected by radial velocity, but 
microlensing will provide true masses for these objects, with no 
$\sin I$ degeneracy. A more comprehensive study of each 
source--lens pair would be needed to calculate the microlensing signatures 
of planets of different masses, at different orbital separations and phases, 
and to evaluate their detectability.

\emph{Gaia}~DR3 is scheduled to be released in 
late 2020, and will extend the time baseline of 
the astrometric measurements analysed by $\sim50\%$. 
We can expect improvements to the 
formal errors on parallax and proper motion 
for our sources and lenses, together with 
the promotion of many objects from 2-parameter 
to 5-parameter astrometric solutions. This will 
improve our estimates for the alignments 
of lenses DR2~4063191108246057472 in May 2021 and 
DR2~4111740078488465408 in the second half of 2033; 
unfortunately, DR2~4116504399886241792 will lens its 
2-parameter background star before DR3, as we discussed above. 
A comparison of the two panels of Figure~\ref{fig:radec} 
shows the improvement in source star motion going from a 
2-parameter astrometric solution (for source 
DR2~4116504399831319424) to a high-quality 5-parameter 
solution (for source DR2~5918299908365843840). The quality 
of the solution for DR2~5918299908365843840 is very good compared 
to most of of source stars, which tend to show a noticeable 
fanning-out of trajectories over the 20-year timescale 
we have considered, as the signal-to-noise on the 
sources' proper motions is often poor. Furthermore, 
\cite{BramichNielsen18} note that our lenses tend to 
have high reduced chi-squared for the \emph{Gaia} astrometric fits. 
This, however, is simply the result of the stars being very red 
($G_\mathrm{BP}-G_\mathrm{RP} \sim 2.9$ to $4.3$): 
compared with stars in general of the same 
magnitude and colour, they are quite normal.
Additionally, many of the lenses can be matched to known 
high-proper motion stars, so these solutions are likely not 
spurious, although they will be improved with DR3. DR3 will 
permit both a refinement on the lensing probabilities 
and magnifications, and allow us to predict 
future events on still longer timescales.

DR3 will also give a more complete and reliable catalogue of 
nearby lens stars. Around 1 in 6 high-proper 
motion stars are missing from \emph{Gaia} DR2 
\citep{GaiaSummary18}, so we can expect another $\sim5$ 
alignments within 0.1 arcsec when DR3 is released 
from the more complete catalogue. 
It will also contain a treatment of 
astrometric binaries, improving the reliability 
of the proper motions of these objects which are 
currently modelled as single stars.

Finally, DR3 will offer improved photometry, 
in particular full spectrophotometry from the 
red and blue photometers and not only 
$G_\mathrm{RP}$ and $G_\mathrm{BP}$ fluxes 
\citep{Andrae+18}. Given the cool temperatures 
of the lenses we have identified, of particular 
interest for lensing studies will be the treatment 
of cool dwarfs \citep[e.g.][]{Sarro+13}. While 
the lenses we have identified from DR2 can be followed 
up and characterised from the ground prior to DR3, 
DR3 will permit better estimates to be made of the 
lens mass when searching for the new alignments 
that DR3 will reveal.

\section{Conclusions}

\begin{itemize}
\item We have searched \emph{Gaia} DR2 for 
  future photometric microlensing events before 2035.5 
  where the lens is within 100\,pc. We have 
  identified 30 source--lens pairs where the 
  nominal proper motion and parallax create 
  an alignment within 0.1 arcsec.
\item Of these 30, 7 pairs have a $>10\%$ 
  probability of passing within one Einstein 
  radius, a regime ideal for detecting 
  low-mass planets orbiting the lens star.
\item Our candidate lens with highest net magnification 
  is DR2~5918299904067162240 
  (WISE J175839.20--583931.6), which will have its 
  closest approach to DR2~5918299908365843840 
  around 1 Apr 2030, with a peak magnification of 
  $23\%$.
\item The first of our seven high-probability lenses 
  is DR2~4116504399886241792 (2MASS J17392440--2327071), 
  which will have its closest approach to the background 
  source in early--mid July 2020. As this predates the 
  improved astrometry expected from \emph{Gaia} DR3 
  in late 2020, further monitoring of this object 
  is warranted to refine the lensing prediction.
\item The Einstein radii of these lenses 
  range from $0.1$ to $0.4$\,au, meaning that 
  any planets found by lensing would fill a region 
  of parameter space around low-mass stars that 
  lies between current transit detections at 
  $<0.1$\,au and current microlensing detections 
  at $\sim1$\,au.
\end{itemize}

Jupyter notebooks to reproduce our analysis, and 
a machine-readable table of results, are available at 
\url{https://github.com/AJMustill/gaia_microlensing}.


\begin{acknowledgements}
The authors are supported by the IMPACT grant 
from the Knut \& Alice Wallenberg Foundation, 
and by the Swedish National Space Board. 
AJM thanks Alcione Mora of the \emph{Gaia} helpdesk for 
clarifications, Paul McMillan for useful discussions, 
the anonymous referee for suggesting a closer look 
at the properties of detectable planets, and Fr\'ed\'eric Th\'evenin 
for drawing our attention to previous work on $\alpha$~Cen. 
This work has made use of data from the European Space Agency (ESA) mission
{\it Gaia} (\url{https://www.cosmos.esa.int/gaia}), processed by the {\it Gaia}
Data Processing and Analysis Consortium (DPAC,
\url{https://www.cosmos.esa.int/web/gaia/dpac/consortium}). Funding for the DPAC
has been provided by national institutions, in particular the institutions
participating in the {\it Gaia} Multilateral Agreement. 
This research has made use of the SIMBAD database 
and the VizieR catalogue tool, operated at CDS, 
Strasbourg, France \citep{Ochsenbein+00,Wenger+00}. 
This research made use of Astropy, a community-developed core 
Python package for Astronomy \citep{Astropy2018}. 
This research has made use of the Extrasolar 
Planets Encyclopaedia (\url{http://exoplanet.eu/}).
\end{acknowledgements}

%
   \bibliographystyle{aa} 
   \bibliography{gaia-microlensing} 

\begin{thebibliography}{42}
\expandafter\ifx\csname natexlab\endcsname\relax\def\natexlab#1{#1}\fi

\bibitem[{{Andrae} {et~al.}(2018){Andrae}, {Fouesneau}, {Creevey}, {Ordenovic},
  {Mary}, {Burlacu}, {Chaoul}, {Jean-Antoine-Piccolo}, {Kordopatis}, {Korn},
  {Lebreton}, {Panem}, {Pichon}, {Thevenin}, {Walmsley}, \& {Bailer-
  Jones}}]{Andrae+18}
{Andrae}, R., {Fouesneau}, M., {Creevey}, O., {et~al.} 2018, ArXiv e-prints,
  arXiv:1804.09374

\bibitem[{{Bond} {et~al.}(2001){Bond}, {Abe}, {Dodd}, {Hearnshaw}, {Honda},
  {Jugaku}, {Kilmartin}, {Marles}, {Masuda}, {Matsubara}, {Muraki}, {Nakamura},
  {Nankivell}, {Noda}, {Noguchi}, {Ohnishi}, {Rattenbury}, {Reid}, {Saito},
  {Sato}, {Sekiguchi}, {Skuljan}, {Sullivan}, {Sumi}, {Takeuti}, {Watase},
  {Wilkinson}, {Yamada}, {Yanagisawa}, \& {Yock}}]{Bond+01}
{Bond}, I.~A., {Abe}, F., {Dodd}, R.~J., {et~al.} 2001, \mnras, 327, 868

\bibitem[{{Bond} {et~al.}(2004){Bond}, {Udalski}, {Jaroszy{\'n}ski},
  {Rattenbury}, {Paczy{\'n}ski}, {Soszy{\'n}ski}, {Wyrzykowski},
  {Szyma{\'n}ski}, {Kubiak}, {Szewczyk}, {{\.Z}ebru{\'n}}, {Pietrzy{\'n}ski},
  {Abe}, {Bennett}, {Eguchi}, {Furuta}, {Hearnshaw}, {Kamiya}, {Kilmartin},
  {Kurata}, {Masuda}, {Matsubara}, {Muraki}, {Noda}, {Okajima}, {Sako},
  {Sekiguchi}, {Sullivan}, {Sumi}, {Tristram}, {Yanagisawa}, {Yock}, \& {OGLE
  Collaboration}}]{Bond+04}
{Bond}, I.~A., {Udalski}, A., {Jaroszy{\'n}ski}, M., {et~al.} 2004, \apj, 606,
  L155

\bibitem[{{Bramich}(2018)}]{Bramich18}
{Bramich}, D.~M. 2018, ArXiv e-prints [\eprint[arXiv]{1805.10630}]

\bibitem[{{Bramich} \& {Nielsen}(2018)}]{BramichNielsen18}
{Bramich}, D.~M. \& {Nielsen}, M.~B. 2018, ArXiv e-prints
  [\eprint[arXiv]{1806.10003}]

\bibitem[{{Catala}(2009)}]{Catala09}
{Catala}, C. 2009, Experimental Astronomy, 23, 329

\bibitem[{{Di Stefano}(2008)}]{DiStefano08}
{Di Stefano}, R. 2008, \apj, 684, 59

\bibitem[{{Di Stefano} {et~al.}(2013){Di Stefano}, {Matthews}, \&
  {L{\'e}pine}}]{DiStefano+13}
{Di Stefano}, R., {Matthews}, J., \& {L{\'e}pine}, S. 2013, \apj, 771, 79

\bibitem[{{Dominik}(1999)}]{Dominik99}
{Dominik}, M. 1999, \aap, 349, 108

\bibitem[{{Einstein}(1936)}]{Einstein36}
{Einstein}, A. 1936, Science, 84, 506

\bibitem[{{Gaia Collaboration} {et~al.}(2018){Gaia Collaboration}, {Brown},
  {Vallenari}, {Prusti}, {de Bruijne}, {Babusiaux}, \&
  {Bailer-Jones}}]{GaiaSummary18}
{Gaia Collaboration}, {Brown}, A.~G.~A., {Vallenari}, A., {et~al.} 2018, ArXiv
  e-prints, arXiv:1804.09365

\bibitem[{{Gaia Collaboration} {et~al.}(2016){Gaia Collaboration}, {Prusti},
  {de Bruijne}, {Brown}, {Vallenari}, {Babusiaux}, {Bailer-Jones}, {Bastian},
  {Biermann}, {Evans}, {Eyer}, {Jansen}, {Jordi}, {Klioner}, {Lammers},
  {Lindegren}, {Luri}, {Mignard}, {Milligan}, {Panem}, {Poinsignon},
  {Pourbaix}, {Randich}, {Sarri}, {Sartoretti}, {Siddiqui}, {Soubiran},
  {Valette}, {van Leeuwen}, {Walton}, {Aerts}, {Arenou}, {Cropper}, {Drimmel},
  {H{\o}g}, {Katz}, {Lattanzi}, {O'Mullane}, {Grebel}, {Holland}, {Huc},
  {Passot}, {Bramante}, {Cacciari}, {Casta{\~n}eda}, {Chaoul}, {Cheek}, {De
  Angeli}, {Fabricius}, {Guerra}, {Hern{\'a}ndez}, {Jean-Antoine-Piccolo},
  {Masana}, {Messineo}, {Mowlavi}, {Nienartowicz}, {Ord{\'o}{\~n}ez- Blanco},
  {Panuzzo}, {Portell}, {Richards}, {Riello}, {Seabroke}, {Tanga},
  {Th{\'e}venin}, {Torra}, {Els}, {Gracia- Abril}, {Comoretto},
  {Garcia-Reinaldos}, {Lock}, {Mercier}, {Altmann}, {Andrae}, {Astraatmadja},
  {Bellas-Velidis}, {Benson}, {Berthier}, {Blomme}, {Busso}, {Carry},
  {Cellino}, {Clementini}, {Cowell}, {Creevey}, {Cuypers}, {Davidson}, {De
  Ridder}, {de Torres}, {Delchambre}, {Dell'Oro}, {Ducourant}, {Fr{\'e}mat},
  {Garc{\'\i}a-Torres}, {Gosset}, {Halbwachs}, {Hambly}, {Harrison}, {Hauser},
  {Hestroffer}, {Hodgkin}, {Huckle}, {Hutton}, {Jasniewicz}, {Jordan},
  {Kontizas}, {Korn}, {Lanzafame}, {Manteiga}, {Moitinho}, {Muinonen},
  {Osinde}, {Pancino}, {Pauwels}, {Petit}, {Recio-Blanco}, {Robin}, {Sarro},
  {Siopis}, {Smith}, {Smith}, {Sozzetti}, {Thuillot}, {van Reeven}, {Viala},
  {Abbas}, {Abreu Aramburu}, {Accart}, {Aguado}, {Allan}, {Allasia},
  {Altavilla}, {{\'A}lvarez}, {Alves}, {Anderson}, {Andrei}, {Anglada Varela},
  {Antiche}, {Antoja}, {Ant{\'o}n}, {Arcay}, {Atzei}, {Ayache}, {Bach},
  {Baker}, {Balaguer-N{\'u}{\~n}ez}, {Barache}, {Barata}, {Barbier}, {Barblan},
  {Baroni}, {Barrado y Navascu{\'e}s}, {Barros}, {Barstow}, {Becciani},
  {Bellazzini}, {Bellei}, {Bello Garc{\'\i}a}, {Belokurov}, {Bendjoya},
  {Berihuete}, {Bianchi}, {Bienaym{\'e}}, {Billebaud}, {Blagorodnova},
  {Blanco-Cuaresma}, {Boch}, {Bombrun}, {Borrachero}, {Bouquillon}, {Bourda},
  {Bouy}, {Bragaglia}, {Breddels}, {Brouillet}, {Br{\"u}semeister},
  {Bucciarelli}, {Budnik}, {Burgess}, {Burgon}, {Burlacu}, {Busonero}, {Buzzi},
  {Caffau}, {Cambras}, {Campbell}, {Cancelliere}, {Cantat-Gaudin}, {Carlucci},
  {Carrasco}, {Castellani}, {Charlot}, {Charnas}, {Charvet}, {Chassat},
  {Chiavassa}, {Clotet}, {Cocozza}, {Collins}, {Collins}, {Costigan}, {Crifo},
  {Cross}, {Crosta}, {Crowley}, {Dafonte}, {Damerdji}, {Dapergolas}, {David},
  {David}, {De Cat}, {de Felice}, {de Laverny}, {De Luise}, {De March}, {de
  Martino}, {de Souza}, {Debosscher}, {del Pozo}, {Delbo}, {Delgado},
  {Delgado}, {di Marco}, {Di Matteo}, {Diakite}, {Distefano}, {Dolding}, {Dos
  Anjos}, {Drazinos}, {Dur{\'a}n}, {Dzigan}, {Ecale}, {Edvardsson}, {Enke},
  {Erdmann}, {Escolar}, {Espina}, {Evans}, {Eynard Bontemps}, {Fabre},
  {Fabrizio}, {Faigler}, {Falc{\~a}o}, {Farr{\`a}s Casas}, {Faye}, {Federici},
  {Fedorets}, {Fern{\'a}ndez-Hern{\'a}ndez}, {Fernique}, {Fienga}, {Figueras},
  {Filippi}, {Findeisen}, {Fonti}, {Fouesneau}, {Fraile}, {Fraser}, {Fuchs},
  {Furnell}, {Gai}, {Galleti}, {Galluccio}, {Garabato}, {Garc{\'\i}a-Sedano},
  {Gar{\'e}}, {Garofalo}, {Garralda}, {Gavras}, {Gerssen}, {Geyer}, {Gilmore},
  {Girona}, {Giuffrida}, {Gomes}, {Gonz{\'a}lez-Marcos},
  {Gonz{\'a}lez-N{\'u}{\~n}ez}, {Gonz{\'a}lez-Vidal}, {Granvik}, {Guerrier},
  {Guillout}, {Guiraud}, {G{\'u}rpide}, {Guti{\'e}rrez-S{\'a}nchez}, {Guy},
  {Haigron}, {Hatzidimitriou}, {Haywood}, {Heiter}, {Helmi}, {Hobbs},
  {Hofmann}, {Holl}, {Holland}, {Hunt}, {Hypki}, {Icardi}, {Irwin}, {Jevardat
  de Fombelle}, {Jofr{\'e}}, {Jonker}, {Jorissen}, {Julbe}, {Karampelas},
  {Kochoska}, {Kohley}, {Kolenberg}, {Kontizas}, {Koposov}, {Kordopatis},
  {Koubsky}, {Kowalczyk}, {Krone-Martins}, {Kudryashova}, {Kull}, {Bachchan},
  {Lacoste-Seris}, {Lanza}, {Lavigne}, {Le Poncin-Lafitte}, {Lebreton},
  {Lebzelter}, {Leccia}, {Leclerc}, {Lecoeur-Taibi}, {Lemaitre}, {Lenhardt},
  {Leroux}, {Liao}, {Licata}, {Lindstr{\o}m}, {Lister}, {Livanou}, {Lobel},
  {L{\"o}ffler}, {L{\'o}pez}, {Lopez-Lozano}, {Lorenz}, {Loureiro},
  {MacDonald}, {Magalh{\~a}es Fernandes}, {Managau}, {Mann}, {Mantelet},
  {Marchal}, {Marchant}, {Marconi}, {Marie}, {Marinoni}, {Marrese},
  {Marschalk{\'o}}, {Marshall}, {Mart{\'\i}n-Fleitas}, {Martino}, {Mary},
  {Matijevi{\v{c}}}, {Mazeh}, {McMillan}, {Messina}, {Mestre}, {Michalik},
  {Millar}, {Miranda}, {Molina}, {Molinaro}, {Molinaro}, {Moln{\'a}r},
  {Moniez}, {Montegriffo}, {Monteiro}, {Mor}, {Mora}, {Morbidelli}, {Morel},
  {Morgenthaler}, {Morley}, {Morris}, {Mulone}, {Muraveva}, {Musella},
  {Narbonne}, {Nelemans}, {Nicastro}, {Noval}, {Ord{\'e}novic},
  {Ordieres-Mer{\'e}}, {Osborne}, {Pagani}, {Pagano}, {Pailler}, {Palacin},
  {Palaversa}, {Parsons}, {Paulsen}, {Pecoraro}, {Pedrosa}, {Pentik{\"a}inen},
  {Pereira}, {Pichon}, {Piersimoni}, {Pineau}, {Plachy}, {Plum}, {Poujoulet},
  {Pr{\v{s}}a}, {Pulone}, {Ragaini}, {Rago}, {Rambaux}, {Ramos-Lerate},
  {Ranalli}, {Rauw}, {Read}, {Regibo}, {Renk}, {Reyl{\'e}}, {Ribeiro},
  {Rimoldini}, {Ripepi}, {Riva}, {Rixon}, {Roelens}, {Romero-G{\'o}mez},
  {Rowell}, {Royer}, {Rudolph}, {Ruiz-Dern}, {Sadowski}, {Sagrist{\`a}
  Sell{\'e}s}, {Sahlmann}, {Salgado}, {Salguero}, {Sarasso}, {Savietto},
  {Schnorhk}, {Schultheis}, {Sciacca}, {Segol}, {Segovia}, {Segransan},
  {Serpell}, {Shih}, {Smareglia}, {Smart}, {Smith}, {Solano}, {Solitro},
  {Sordo}, {Soria Nieto}, {Souchay}, {Spagna}, {Spoto}, {Stampa}, {Steele},
  {Steidelm{\"u}ller}, {Stephenson}, {Stoev}, {Suess}, {S{\"u}veges}, {Surdej},
  {Szabados}, {Szegedi-Elek}, {Tapiador}, {Taris}, {Tauran}, {Taylor},
  {Teixeira}, {Terrett}, {Tingley}, {Trager}, {Turon}, {Ulla}, {Utrilla},
  {Valentini}, {van Elteren}, {Van Hemelryck}, {van Leeuwen}, {Varadi},
  {Vecchiato}, {Veljanoski}, {Via}, {Vicente}, {Vogt}, {Voss}, {Votruba},
  {Voutsinas}, {Walmsley}, {Weiler}, {Weingrill}, {Werner}, {Wevers},
  {Whitehead}, {Wyrzykowski}, {Yoldas}, {{\v{Z}}erjal}, {Zucker}, {Zurbach},
  {Zwitter}, {Alecu}, {Allen}, {Allende Prieto}, {Amorim},
  {Anglada-Escud{\'e}}, {Arsenijevic}, {Azaz}, {Balm}, {Beck}, {Bernstein},
  {Bigot}, {Bijaoui}, {Blasco}, {Bonfigli}, {Bono}, {Boudreault}, {Bressan},
  {Brown}, {Brunet}, {Bunclark}, {Buonanno}, {Butkevich}, {Carret}, {Carrion},
  {Chemin}, {Ch{\'e}reau}, {Corcione}, {Darmigny}, {de Boer}, {de Teodoro}, {de
  Zeeuw}, {Delle Luche}, {Domingues}, {Dubath}, {Fodor}, {Fr{\'e}zouls},
  {Fries}, {Fustes}, {Fyfe}, {Gallardo}, {Gallegos}, {Gardiol}, {Gebran},
  {Gomboc}, {G{\'o}mez}, {Grux}, {Gueguen}, {Heyrovsky}, {Hoar}, {Iannicola},
  {Isasi Parache}, {Janotto}, {Joliet}, {Jonckheere}, {Keil}, {Kim},
  {Klagyivik}, {Klar}, {Knude}, {Kochukhov}, {Kolka}, {Kos}, {Kutka}, {Lainey},
  {LeBouquin}, {Liu}, {Loreggia}, {Makarov}, {Marseille}, {Martayan},
  {Martinez-Rubi}, {Massart}, {Meynadier}, {Mignot}, {Munari}, {Nguyen},
  {Nordlander}, {Ocvirk}, {O'Flaherty}, {Olias Sanz}, {Ortiz}, {Osorio},
  {Oszkiewicz}, {Ouzounis}, {Palmer}, {Park}, {Pasquato}, {Peltzer}, {Peralta},
  {P{\'e}turaud}, {Pieniluoma}, {Pigozzi}, {Poels}, {Prat}, {Prod'homme},
  {Raison}, {Rebordao}, {Risquez}, {Rocca-Volmerange}, {Rosen}, {Ruiz-Fuertes},
  {Russo}, {Sembay}, {Serraller Vizcaino}, {Short}, {Siebert}, {Silva},
  {Sinachopoulos}, {Slezak}, {Soffel}, {Sosnowska}, {Strai{\v{z}}ys}, {ter
  Linden}, {Terrell}, {Theil}, {Tiede}, {Troisi}, {Tsalmantza}, {Tur},
  {Vaccari}, {Vachier}, {Valles}, {Van Hamme}, {Veltz}, {Virtanen}, {Wallut},
  {Wichmann}, {Wilkinson}, {Ziaeepour}, \& {Zschocke}}]{GaiaMission16}
{Gaia Collaboration}, {Prusti}, T., {de Bruijne}, J.~H.~J., {et~al.} 2016,
  \aap, 595, A1

\bibitem[{{Gaudi}(2012)}]{Gaudi12}
{Gaudi}, B.~S. 2012, Annual Review of Astronomy and Astrophysics, 50, 411

\bibitem[{{Giorgini} {et~al.}(1996){Giorgini}, {Yeomans}, {Chamberlin},
  {Chodas}, {Jacobson}, {Keesey}, {Lieske}, {Ostro}, {Standish}, \&
  {Wimberly}}]{Giorgini+96}
{Giorgini}, J.~D., {Yeomans}, D.~K., {Chamberlin}, A.~B., {et~al.} 1996, in
  American Astronomical Society, DPS meeting 28, id.25.04, Vol.~28, 25.04

\bibitem[{{Harding} {et~al.}(2018){Harding}, {Stefano}, {L{\'e}pine}, {Urama},
  {Pham}, \& {Baker}}]{Harding+18}
{Harding}, A.~J., {Stefano}, R.~D., {L{\'e}pine}, S., {et~al.} 2018, \mnras,
  475, 79

\bibitem[{{Hwang} {et~al.}(2018){Hwang}, {Ryu}, {Kim}, {Albrow}, {Chung},
  {Gould}, {Han}, {Jung}, {Shin}, {Shvartzvald}, {Yee}, {Zang}, {Cha}, {Kim},
  {Kim}, {Lee}, {Lee}, {Lee}, {Park}, \& {Pogge}}]{Hwang+18}
{Hwang}, K.-H., {Ryu}, Y.-H., {Kim}, H.-W., {et~al.} 2018, ArXiv e-prints,
  arXiv:1805.08888

\bibitem[{{Jung} {et~al.}(2018{\natexlab{a}}){Jung}, {Hwang}, {Ryu}, {Gould},
  {Han}, {Yee}, {Albrow}, {Chung}, {Shin}, {Shvartzvald}, {Zang}, {Cha}, {Kim},
  {Kim}, {Kim}, {Lee}, {Lee}, {Lee}, {Park}, \& {Pogge}}]{Jung+18b}
{Jung}, Y.~K., {Hwang}, K.-H., {Ryu}, Y.-H., {et~al.} 2018{\natexlab{a}}, ArXiv
  e-prints [\eprint[arXiv]{1805.09983}]

\bibitem[{{Jung} {et~al.}(2018{\natexlab{b}}){Jung}, {Udalski}, {Gould}, {Ryu},
  {Yee}, {and}, {Han}, {Albrow}, {Lee}, {Kim}, {Hwang}, {Chung}, {Shin}, {Zhu},
  {Cha}, {Kim}, {Lee}, {Park}, {Lee}, {Kim}, {Pogge}, {The KMTNet
  Collaboration}, {Szyma{\'n}ski}, {Mr{\'o}z}, {Poleski}, {Skowron},
  {Pietrukowicz}, {Soszy{\'n}ski}, {Koz{\l}owski}, {Ulaczyk}, {Pawlak},
  {Rybicki}, \& {The OGLE Collaboration}}]{Jung+18a}
{Jung}, Y.~K., {Udalski}, A., {Gould}, A., {et~al.} 2018{\natexlab{b}}, \aj,
  155, 219

\bibitem[{{Kervella} {et~al.}(2016){Kervella}, {Mignard}, {M{\'e}rand}, \&
  {Th{\'e}venin}}]{Kervella+16}
{Kervella}, P., {Mignard}, F., {M{\'e}rand}, A., \& {Th{\'e}venin}, F. 2016,
  \aap, 594, A107

\bibitem[{{Kim} {et~al.}(2016){Kim}, {Lee}, {Park}, {Kim}, {Cha}, {Lee}, {Han},
  {Chun}, \& {Yuk}}]{Kim+16}
{Kim}, S.-L., {Lee}, C.-U., {Park}, B.-G., {et~al.} 2016, Journal of Korean
  Astronomical Society, 49, 37

\bibitem[{{Kl{\"u}ter} {et~al.}(2018){Kl{\"u}ter}, {Bastian}, {Demleitner}, \&
  {Wambsganss}}]{Klueter+18}
{Kl{\"u}ter}, J., {Bastian}, U., {Demleitner}, M., \& {Wambsganss}, J. 2018,
  ArXiv e-prints [\eprint[arXiv]{1805.08023}]

\bibitem[{{L{\'e}pine}(2008)}]{Lepine08}
{L{\'e}pine}, S. 2008, \aj, 135, 2177

\bibitem[{{L{\'e}pine} {et~al.}(2003){L{\'e}pine}, {Rich}, \&
  {Shara}}]{Lepine+03}
{L{\'e}pine}, S., {Rich}, R.~M., \& {Shara}, M.~M. 2003, \aj, 125, 1598

\bibitem[{{L{\'e}pine} \& {Shara}(2005)}]{LepineShara05}
{L{\'e}pine}, S. \& {Shara}, M.~M. 2005, \aj, 129, 1483

\bibitem[{{Lindegren} {et~al.}(2018){Lindegren}, {Hernandez}, {Bombrun},
  {Klioner}, {Bastian}, {Ramos-Lerate}, {de Torres}, {Steidelmuller},
  {Stephenson}, {Hobbs}, {Lammers}, {Biermann}, {Geyer}, {Hilger}, {Michalik},
  {Stampa}, {McMillan}, {Castaneda}, {Clotet}, {Comoretto}, {Davidson},
  {Fabricius}, {Gracia}, {Hambly}, {Hutton}, {Mora}, {Portell}, {van Leeuwen},
  {Abbas}, {Abreu}, {Altmann}, {Andrei}, {Anglada}, {Balaguer- Nunez},
  {Barache}, {Becciani}, {Bertone}, {Bianchi}, {Bouquillon}, {Bourda},
  {Brusemeister}, {Bucciarelli}, {Busonero}, {Buzzi}, {Cancelliere},
  {Carlucci}, {Charlot}, {Cheek}, {Crosta}, {Crowley}, {de Bruijne}, {de
  Felice}, {Drimmel}, {Esquej}, {Fienga}, {Fraile}, {Gai}, {Garralda},
  {Gonzalez-Vidal}, {Guerra}, {Hauser}, {Hofmann}, {Holl}, {Jordan},
  {Lattanzi}, {Lenhardt}, {Liao}, {Licata}, {Lister}, {Loffler}, {Marchant},
  {Martin-Fleitas}, {Messineo}, {Mignard}, {Morbidelli}, {Poggio}, {Riva},
  {Rowell}, {Salguero}, {Sarasso}, {Sciacca}, {Siddiqui}, {Smart}, {Spagna},
  {Steele}, {Taris}, {Torra}, {van Elteren}, {van Reeven}, \&
  {Vecchiato}}]{Lindegren+18}
{Lindegren}, L., {Hernandez}, J., {Bombrun}, A., {et~al.} 2018, ArXiv e-prints,
  arXiv:1804.09366

\bibitem[{{Luhman} \& {Sheppard}(2014)}]{LuhmanSheppard14}
{Luhman}, K.~L. \& {Sheppard}, S.~S. 2014, \apj, 787, 126

\bibitem[{{Luri} {et~al.}(2018){Luri}, {Brown}, {Sarro}, {Arenou},
  {Bailer-Jones}, {Castro-Ginard}, {de Bruijne}, {Prusti}, {Babusiaux}, \&
  {Delgado}}]{Luri+18}
{Luri}, X., {Brown}, A.~G.~A., {Sarro}, L.~M., {et~al.} 2018, ArXiv e-prints,
  arXiv:1804.09376

\bibitem[{{McGill} {et~al.}(2018){McGill}, {Smith}, {Wyn Evans}, {Belokurov},
  \& {Smart}}]{McGill+18}
{McGill}, P., {Smith}, L.~C., {Wyn Evans}, N., {Belokurov}, V., \& {Smart},
  R.~L. 2018, \mnras, L72

\bibitem[{{Ochsenbein} {et~al.}(2000){Ochsenbein}, {Bauer}, \&
  {Marcout}}]{Ochsenbein+00}
{Ochsenbein}, F., {Bauer}, P., \& {Marcout}, J. 2000, Astronomy and
  Astrophysics Supplement Series, 143, 23

\bibitem[{{Paczynski}(1986)}]{Paczynski86a}
{Paczynski}, B. 1986, \apj, 301, 503

\bibitem[{{Paczynski}(1995)}]{Paczynski95}
{Paczynski}, B. 1995, \actaa, 45, 345

\bibitem[{{Paczynski}(1996)}]{Paczynski96}
{Paczynski}, B. 1996, Annual Review of Astronomy and Astrophysics, 34, 419

\bibitem[{{Paczynski}(2001)}]{Paczynski01}
{Paczynski}, B. 2001, ArXiv e-prints, astro

\bibitem[{{Rauer} {et~al.}(2014){Rauer}, {Catala}, {Aerts}, {Appourchaux},
  {Benz}, {Brandeker}, {Christensen-Dalsgaard}, {Deleuil}, {Gizon}, {Goupil},
  {G{\"u}del}, {Janot-Pacheco}, {Mas-Hesse}, {Pagano}, {Piotto}, {Pollacco},
  {Santos}, {Smith}, {Su{\'a}rez}, {Szab{\'o}}, {Udry}, {Adibekyan}, {Alibert},
  {Almenara}, {Amaro-Seoane}, {Eiff}, {Asplund}, {Antonello}, {Barnes},
  {Baudin}, {Belkacem}, {Bergemann}, {Bihain}, {Birch}, {Bonfils}, {Boisse},
  {Bonomo}, {Borsa}, {Brand{\~a}o}, {Brocato}, {Brun}, {Burleigh}, {Burston},
  {Cabrera}, {Cassisi}, {Chaplin}, {Charpinet}, {Chiappini}, {Church},
  {Csizmadia}, {Cunha}, {Damasso}, {Davies}, {Deeg}, {D{\'\i}az}, {Dreizler},
  {Dreyer}, {Eggenberger}, {Ehrenreich}, {Eigm{\"u}ller}, {Erikson}, {Farmer},
  {Feltzing}, {de Oliveira Fialho}, {Figueira}, {Forveille}, {Fridlund},
  {Garc{\'\i}a}, {Giommi}, {Giuffrida}, {Godolt}, {Gomes da Silva}, {Granzer},
  {Grenfell}, {Grotsch-Noels}, {G{\"u}nther}, {Haswell}, {Hatzes},
  {H{\'e}brard}, {Hekker}, {Helled}, {Heng}, {Jenkins}, {Johansen},
  {Khodachenko}, {Kislyakova}, {Kley}, {Kolb}, {Krivova}, {Kupka}, {Lammer},
  {Lanza}, {Lebreton}, {Magrin}, {Marcos-Arenal}, {Marrese}, {Marques},
  {Martins}, {Mathis}, {Mathur}, {Messina}, {Miglio}, {Montalban}, {Montalto},
  {Monteiro}, {Moradi}, {Moravveji}, {Mordasini}, {Morel}, {Mortier},
  {Nascimbeni}, {Nelson}, {Nielsen}, {Noack}, {Norton}, {Ofir}, {Oshagh},
  {Ouazzani}, {P{\'a}pics}, {Parro}, {Petit}, {Plez}, {Poretti}, {Quirrenbach},
  {Ragazzoni}, {Raimondo}, {Rainer}, {Reese}, {Redmer}, {Reffert},
  {Rojas-Ayala}, {Roxburgh}, {Salmon}, {Santerne}, {Schneider}, {Schou},
  {Schuh}, {Schunker}, {Silva-Valio}, {Silvotti}, {Skillen}, {Snellen}, {Sohl},
  {Sousa}, {Sozzetti}, {Stello}, {Strassmeier}, {{\v{S}}vanda}, {Szab{\'o}},
  {Tkachenko}, {Valencia}, {Van Grootel}, {Vauclair}, {Ventura}, {Wagner},
  {Walton}, {Weingrill}, {Werner}, {Wheatley}, \& {Zwintz}}]{Rauer+14}
{Rauer}, H., {Catala}, C., {Aerts}, C., {et~al.} 2014, Experimental Astronomy,
  38, 249

\bibitem[{{Ricker} {et~al.}(2015){Ricker}, {Winn}, {Vanderspek}, {Latham},
  {Bakos}, {Bean}, {Berta-Thompson}, {Brown}, {Buchhave}, {Butler}, {Butler},
  {Chaplin}, {Charbonneau}, {Christensen-Dalsgaard}, {Clampin}, {Deming},
  {Doty}, {De Lee}, {Dressing}, {Dunham}, {Endl}, {Fressin}, {Ge}, {Henning},
  {Holman}, {Howard}, {Ida}, {Jenkins}, {Jernigan}, {Johnson}, {Kaltenegger},
  {Kawai}, {Kjeldsen}, {Laughlin}, {Levine}, {Lin}, {Lissauer}, {MacQueen},
  {Marcy}, {McCullough}, {Morton}, {Narita}, {Paegert}, {Palle}, {Pepe},
  {Pepper}, {Quirrenbach}, {Rinehart}, {Sasselov}, {Sato}, {Seager},
  {Sozzetti}, {Stassun}, {Sullivan}, {Szentgyorgyi}, {Torres}, {Udry}, \&
  {Villasenor}}]{Ricker+15}
{Ricker}, G.~R., {Winn}, J.~N., {Vanderspek}, R., {et~al.} 2015, Journal of
  Astronomical Telescopes, Instruments, and Systems, 1, 014003

\bibitem[{{Salaris} \& {Cassisi}(2005)}]{SalarisCassisi05}
{Salaris}, M. \& {Cassisi}, S. 2005, {Evolution of Stars and Stellar
  Populations}

\bibitem[{{Sarro} {et~al.}(2013){Sarro}, {Berihuete}, {Carri{\'o}n}, {Barrado},
  {Cruz}, \& {Isasi}}]{Sarro+13}
{Sarro}, L.~M., {Berihuete}, A., {Carri{\'o}n}, C., {et~al.} 2013, \aap, 550,
  A44

\bibitem[{{Shvartzvald} {et~al.}(2016){Shvartzvald}, {Maoz}, {Udalski}, {Sumi},
  {Friedmann}, {Kaspi}, {Poleski}, {Szyma{\'n}ski}, {Skowron}, {Koz{\l}owski},
  {Wyrzykowski}, {Mr{\'o}z}, {Pietrukowicz}, {Pietrzy{\'n}ski},
  {Soszy{\'n}ski}, {Ulaczyk}, {Abe}, {Barry}, {Bennett}, {Bhattacharya},
  {Bond}, {Freeman}, {Inayama}, {Itow}, {Koshimoto}, {Ling}, {Masuda}, {Fukui},
  {Matsubara}, {Muraki}, {Ohnishi}, {Rattenbury}, {Saito}, {Sullivan},
  {Suzuki}, {Tristram}, {Wakiyama}, \& {Yonehara}}]{Shvartzvald+16}
{Shvartzvald}, Y., {Maoz}, D., {Udalski}, A., {et~al.} 2016, \mnras, 457, 4089

\bibitem[{{Suzuki} {et~al.}(2016){Suzuki}, {Bennett}, {Sumi}, {Bond}, {Rogers},
  {Abe}, {Asakura}, {Bhattacharya}, {Donachie}, {Freeman}, {Fukui}, {Hirao},
  {Itow}, {Koshimoto}, {Li}, {Ling}, {Masuda}, {Matsubara}, {Muraki},
  {Nagakane}, {Onishi}, {Oyokawa}, {Rattenbury}, {Saito}, {Sharan}, {Shibai},
  {Sullivan}, {Tristram}, {Yonehara}, \& {MOA Collaboration}}]{Suzuki+16}
{Suzuki}, D., {Bennett}, D.~P., {Sumi}, T., {et~al.} 2016, \apj, 833, 145

\bibitem[{{The Astropy Collaboration} {et~al.}(2018){The Astropy
  Collaboration}, {Price-Whelan}, {Sip{\'{o}}cz}, {G{\"u}nther}, {Lim},
  {Crawford}, {Conseil}, {Shupe}, {Craig}, {Dencheva}, {Ginsburg},
  {VanderPlas}, {Bradley}, {P{\'e}rez- Su{\'a}rez}, {de Val-Borro}, {Aldcroft},
  {Cruz}, {Robitaille}, {Tollerud}, {Ardelean}, {Babej}, {Bachetti}, {Bakanov},
  {Bamford}, {Barentsen}, {Barmby}, {Baumbach}, {Berry}, {Biscani}, {Boquien},
  {Bostroem}, {Bouma}, {Brammer}, {Bray}, {Breytenbach}, {Buddelmeijer},
  {Burke}, {Calderone}, {Cano Rodr{\'\i}guez}, {Cara}, {Cardoso}, {Cheedella},
  {Copin}, {Crichton}, {D{\'A}vella}, {Deil}, {Depagne}, {Dietrich}, {Donath},
  {Droettboom}, {Earl}, {Erben}, {Fabbro}, {Ferreira}, {Finethy}, {Fox},
  {Garrison}, {Gibbons}, {Goldstein}, {Gommers}, {Greco}, {Greenfield},
  {Groener}, {Grollier}, {Hagen}, {Hirst}, {Homeier}, {Horton}, {Hosseinzadeh},
  {Hu}, {Hunkeler}, {Ivezi{\'c}}, {Jain}, {Jenness}, {Kanarek}, {Kendrew},
  {Kern}, {Kerzendorf}, {Khvalko}, {King}, {Kirkby}, {Kulkarni}, {Kumar},
  {Lee}, {Lenz}, {Littlefair}, {Ma}, {Macleod}, {Mastropietro}, {McCully},
  {Montagnac}, {Morris}, {Mueller}, {Mumford}, {Muna}, {Murphy}, {Nelson},
  {Nguyen}, {Ninan}, {N{\"o}the}, {Ogaz}, {Oh}, {Parejko}, {Parley}, {Pascual},
  {Patil}, {Patil}, {Plunkett}, {Prochaska}, {Rastogi}, {Reddy Janga},
  {Sabater}, {Sakurikar}, {Seifert}, {Sherbert}, {Sherwood-Taylor}, {Shih},
  {Sick}, {Silbiger}, {Singanamalla}, {Singer}, {Sladen}, {Sooley},
  {Sornarajah}, {Streicher}, {Teuben}, {Thomas}, {Tremblay}, {Turner},
  {Terr{\'o}n}, {van Kerkwijk}, {de la Vega}, {Watkins}, {Weaver}, {Whitmore},
  {Woillez}, \& {Zabalza}}]{Astropy2018}
{The Astropy Collaboration}, {Price-Whelan}, A.~M., {Sip{\'{o}}cz}, B.~M.,
  {et~al.} 2018, ArXiv e-prints, arXiv:1801.02634

\bibitem[{{Udalski}(2003)}]{Udalski03}
{Udalski}, A. 2003, \actaa, 53, 291

\bibitem[{{Wenger} {et~al.}(2000){Wenger}, {Ochsenbein}, {Egret}, {Dubois},
  {Bonnarel}, {Borde}, {Genova}, {Jasniewicz}, {Lalo{\"e}}, {Lesteven}, \&
  {Monier}}]{Wenger+00}
{Wenger}, M., {Ochsenbein}, F., {Egret}, D., {et~al.} 2000, Astronomy and
  Astrophysics Supplement Series, 143, 9

\end{thebibliography}
%

\begin{appendix}

\section{Details for all alignments}

Table~\ref{tab:all} contains details of all 30 
source--lens pairs whose nominal proper motion and 
parallax takes them within 0.1 arcsec.

\begin{sidewaystable*}
\caption{Details of all 30 source--lens pairs approaching within 0.1\,arcsec, ordered chronoligically by 
  the median epoch of peak magnification. Columns show the \emph{Gaia} DR2 id and $G$ magnitude of the 
  lens and the source, the probability of passing within an Einstein radius, the median 
  and 68\% confidence interval on the net magnification (including dilution by the lens) $f_\mathrm{max}$, 
  the 68\% confidence interval on the epoch of peak magnification $T_\mathrm{max}$, the median and 68\% 
  confidence interval on the duration for which the magnification exceeds $10^{-4}$ (for MC samples in 
  which it does so) $\Delta t$, and the effective temperature and mass of the lens. We provide alternative 
  lens identifiers in the footnotes (marked `?' where not certain).}
\label{tab:all}
\centering
\begin{tabular}{ccccccccccc}
\hline\\
Lens DR2 id & Lens $G$ & Source DR 2 id & Source $G$ & $P(d<1R_\mathrm{E})$ [\%] & $f_\mathrm{max}$ &       $T_\mathrm{max}$ & $\Delta t$ [d] & $T_\mathrm{eff}$ [K] & $M_\star$ [M$_\odot$]\\
\hline\\
4116504399886241792\tablefootmark{a} & 13.86 & 4116504399831319424\tablefootmark{b} & 18.85 & 19.95 & $2.86^{+49.70}_{-2.56}\times10^{-4}$ & 2020 Jul 3--22 & $25^{+4}_{-8}$ & 3554 & 0.16 \\
4063191108246057472 & 13.65 & 4063191073886316928\tablefootmark{b} & 18.86 & 0.00 & $8.89^{+14.14}_{-4.77}\times10^{-7}$ & 2021 Apr 30 -- May 25 & $12^{+5}_{-2}$ & 3900 & 0.19 \\
5901094750438455808 & 16.61 & 5901094746125412608 & 18.97 & 0.00 & $2.34^{+2.70}_{-1.13}\times10^{-4}$ & 2022 Aug 8--19 & $16^{+8}_{-7}$ & 3292 & 0.11 \\
5411468169440274560\tablefootmark{c} & 16.61 & 5411468169435427328 & 20.19 & 6.07 & $8.98^{+135.20}_{-7.40}\times10^{-5}$ & 2022 Aug 2 -- Oct 4 & $62^{+36}_{-22}$ & 3772 & 0.10 \\
5800510224109658240 & 17.72 & 5800510224109658368 & 20.09 & 0.00 & $5.44^{+6.98}_{-2.76}\times10^{-6}$ & 2023 May 2--15 & $12^{+0}_{-0}$ & -- & 0.10\tablefootmark{d} \\
5613074690219333504 & 16.04 & 5613074690220317312 & 20.84 & 0.00 & $3.70^{+4.29}_{-1.81}\times10^{-7}$ & 2024 Sep 27 -- Oct 18 & $0^{+0}_{-0}$ & 3834 & 0.14 \\
3340477717172813568 & 10.10 & 3340477712874873344 & 20.50 & 0.00 & $1.78^{+1.59}_{-0.77}\times10^{-7}$ & 2025 Jul 12--17 & $5^{+4}_{-2}$ & 3428 & 0.26 \\
2042703905028908928\tablefootmark{e} & 14.70 & 2042703900727995008 & 18.60 & 76.77 & $2.13^{+7.59}_{-1.43}\times10^{-2}$ & 2025 Jul 20--27 & $42^{+1}_{-1}$ & 3655 & 0.17 \\
2030898139472914688 & 15.62 & 2030898105088410112 & 19.92 & 49.29 & $6.19^{+36.20}_{-5.23}\times10^{-3}$ & 2025 Sep 1--16 & $30^{+2}_{-4}$ & 4002 & 0.14 \\
5529180853514968960\tablefootmark{f} & 14.77 & 5529180853515235072 & 20.64 & 4.66 & $5.57^{+94.68}_{-4.61}\times10^{-6}$ & 2026 Feb 27 -- Apr 8 & $28^{+5}_{-12}$ & 3922 & 0.18 \\
5903487940560263936\tablefootmark{g} & 14.63 & 5903487940560263808 & 17.86 & 44.60 & $1.34^{+6.31}_{-0.96}\times10^{-2}$ & 2026 May 25 -- Jun 7 & $46^{+3}_{-5}$ & 4065 & 0.14 \\
5819622622435541888 & 16.83 & 5819622622430477184 & 20.22 & 0.00 & $9.19^{+12.81}_{-4.62}\times10^{-7}$ & 2027 Jan 14 -- Feb 13 & $32^{+0}_{-0}$ & 3982 & 0.14 \\
4114040737846721152\tablefootmark{h} & 15.89 & 4114040737803216768 & 20.19 & 7.25 & $2.34^{+81.07}_{-2.06}\times10^{-5}$ & 2027 Mar 9 -- Apr 16 & $38^{+7}_{-15}$ & 3400 & 0.15 \\
4301535507904403712 & 19.09 & 4301535507885645312 & 20.19 & 0.89 & $9.10^{+72.86}_{-6.75}\times10^{-6}$ & 2027 Feb 9 -- May 12 & $34^{+8}_{-15}$ & -- & 0.10\tablefootmark{a} \\
4744453460527256448 & 16.91 & 4744453460527256576 & 16.79 & 0.00 & $4.82^{+0.48}_{-0.43}\times10^{-6}$ & 2028 Feb 5--7 & $0^{+0}_{-0}$ & 3737 & 0.15 \\
318399257231042304\tablefootmark{i} & 17.27 & 318399257231042048 & 17.94 & 35.85 & $5.31^{+45.34}_{-4.58}\times10^{-2}$ & 2028 Jul 10--27 & $54^{+6}_{-16}$ & -- & 0.10\tablefootmark{a} \\
5918299904067162240\tablefootmark{j} & 14.91 & 5918299908365843840 & 13.90 & 45.93 & $2.27^{+2.76}_{-1.11}\times10^{-1}$ & 2030 Mar 31 -- Apr 2 & $14^{+0}_{-1}$ & 3613 & 0.10 \\
2214946451667611904\tablefootmark{k} & 14.55 & 2214946455960518400 & 20.33 & 0.13 & $2.40^{+6.47}_{-1.50}\times10^{-6}$ & 2030 Oct 31 -- Dec 9 & $26^{+12}_{-14}$ & 4140 & 0.30 \\
5191076183243196160\tablefootmark{l} & 16.44 & 5191076183240735232 & 19.84 & 4.27 & $1.43^{+43.64}_{-1.27}\times10^{-5}$ & 2031 Jun 17 -- Aug 28 & $80^{+61}_{-30}$ & 4306 & 0.17 \\
6870168805525824384 & 12.95 & 6870168801226177408 & 18.93 & 0.00 & $1.75^{+1.44}_{-0.71}\times10^{-6}$ & 2033 Jan 30 -- Feb 5 & $0^{+0}_{-0}$ & 3664 & 0.26 \\
5846206202262355712 & 16.99 & 5846206202256330240 & 20.25 & 5.67 & $1.25^{+17.80}_{-1.03}\times10^{-4}$ & 2033 Jan 22 -- Mar 5 & $35^{+11}_{-15}$ & 3296 & 0.09 \\
6663600360559308672 & 10.74 & 6663600291839828096 & 18.00 & 0.00 & $7.97^{+8.47}_{-3.71}\times10^{-6}$ & 2033 Apr 22--23 & $5^{+3}_{-3}$ & 3302 & 0.21 \\
2052808897922015360 & 16.60 & 2052808893622868352 & 18.45 & 0.00 & $1.45^{+1.04}_{-0.57}\times10^{-5}$ & 2033 Jun 18 -- Jul 2 & $0^{+0}_{-0}$ & 3836 & 0.15 \\
116048577524991872\tablefootmark{m} & 15.46 & 116048577524443520 & 20.66 & 4.90 & $2.02^{+118.42}_{-1.81}\times10^{-6}$ & 2033 Jul 23 -- Sep 13 & $37^{+13}_{-13}$ & 3795 & 0.14 \\
4111740078488465408 & 16.38 & 4111740074149205632\tablefootmark{b} & 20.40 & 5.17 & $3.68^{+334.98}_{-3.49}\times10^{-6}$ & 2033 Jun 14 -- 2034 Jan 5 & $36^{+13}_{-13}$ & 4284 & 0.13 \\
5232662289996360448\tablefootmark{n} & 15.57 & 5232662289986673536 & 20.45 & 6.03 & $4.69^{+267.71}_{-4.24}\times10^{-6}$ & 2033 Sep 26 -- 2034 Sep 22 & $68^{+85}_{-22}$ & 3741 & 0.19 \\
2934637564368867072\tablefootmark{o} & 15.49 & 2934637564369093632 & 20.74 & 8.77 & $1.01^{+61.34}_{-0.97}\times10^{-5}$ & 2033 Aug 13 -- 2034 Aug 11 & $58^{+68}_{-20}$ & 3480 & 0.16 \\
3425318817510655360\tablefootmark{p} & 16.15 & 3425318813215052288 & 19.35 & 27.96 & $3.42^{+42.39}_{-3.05}\times10^{-3}$ & 2034 Sep 7--19 & $10^{+1}_{-3}$ & 3546 & 0.15 \\
4612462518451692672 & 16.42 & 4612462514156484864 & 17.56 & 0.00 & $2.51^{+0.51}_{-0.40}\times10^{-6}$ & 2034 Nov 26 -- Dec 5 & $0^{+0}_{-0}$ & 3783 & 0.15 \\
5237644830027339904\tablefootmark{q} & 16.87 & 5237644830015225728 & 19.75 & 0.00 & $4.97^{+4.12}_{-2.04}\times10^{-7}$ & 2035 Jul 2--2\tablefootmark{r} & $0^{+0}_{-0}$ & 3670 & 0.12 \\
\hline
\end{tabular}
\tablefoottext{a}{2MASS J17392440-2327071.}
\tablefoottext{b}{2-parameter astrometric solution only.}
\tablefoottext{c}{USNO-B1.0 0436-0183244?}
\tablefoottext{d}{Assumed.}
\tablefoottext{e}{G 207-14}.
\tablefoottext{f}{USNO-B1.0 0517-0163728?}
\tablefoottext{g}{USNO-B1.0 0421-0506943?}
\tablefoottext{h}{WISE J171513.77-243036.2.}
\tablefoottext{i}{LSPM J0146+3545.}
\tablefoottext{j}{WISE J175839.20-583931.6.}
\tablefoottext{k}{LSPM J2311+6858.}
\tablefoottext{l}{USNO-B1.0 0053-0028550.}
\tablefoottext{m}{LP 354-291.}
\tablefoottext{n}{USNO-B1.0 0212-0238785?}
\tablefoottext{o}{USNO-B1.0 0727-0162208?}
\tablefoottext{p}{2MASS J06095230+2319143.}
\tablefoottext{q}{USNO-B1.0 0259-0216569?}
\tablefoottext{r}{Lens and source have near-parallel proper motion 
  and there is a small ($\sim1\%$) chance of passing within $1R_\mathrm{E}$ 
  in 2036; DR3 will better constrain this.}
\end{sidewaystable*}

\end{appendix}

\end{document}